\documentstyle[prd,aps,preprint,epsfig,tighten]{revtex}
\begin{document}
\preprint{\sf                                  \vbox{\hbox{IASSNS-AST 96/46}   
                                                     \hbox{BARI-TH/248-96}}}
\draft
\title{\Large\bf            Solar neutrinos:                              \\ 
                Near-far asymmetry and just-so oscillations                }
%
\author{	 \ \	B.~Fa{\"\i}d$\,^c$,\ \  
			G.~L.~Fogli$\,^b$,\ \ 
			E.~Lisi$\,^{a,b}$, and\ \ 
			D.~Montanino$\,^b$\\[3mm]}
\address{$^a$Institute for Advanced Study, Princeton, New Jersey 08540\\
         $^b$Dipartimento di Fisica dell'Universit{\`a} and Sezione INFN, 
         I-70126 Bari, Italy  \\
         $^c$Institut de Physique, Universit{\'e} des Sciences et de
	 la Technologie, DZ-16111 Algiers, Algeria}
\maketitle
\begin{abstract}
We propose to study possible signals of just-so oscillations in 
new-generation solar neutrino experiments by separating the events detected 
when the earth is nearest to the sun (perihelion $\pm$~3~months) from those 
detected when the earth is farthest from the sun (aphelion $\pm$~3~months). 
We introduce a solar model independent near-far asymmetry, which is non-zero 
if just-so oscillations occur. We apply our calculations to the kinetic 
energy spectra of electrons induced by $^8$B solar neutrino interactions
 in the SuperKamiokande and Sudbury Neutrino Observatory  experiments. 
We show that the sensitivity to the neutrino oscillation parameters can 
be increased by probing the near-far asymmetry in selected parts of the 
electron energy spectra.
\end{abstract}
\section{INTRODUCTION}

	Neutrino flavor oscillations in vacuum \cite{Po67,Ka62,Bi78} 
represent a viable solution \cite{Gl87,Pe96,Ac91,Ca95,Be96,Ba96} to the 
observed deficit of the solar neutrino flux \cite{Da94,GALL,SAGE,Kami} as 
compared to the predictions of the standard solar model \cite{Ba89,Pi95}.

	If the oscillations have a wavelength comparable to the earth orbit 
radius (just-so oscillations \cite{Gl87}), then significant distortions 
could arise both in the neutrino energy spectrum \cite{Ba69,Ba81,Kr95}
(as a result of the energy-dependence of the oscillation probability) and 
in the time structure of the signal \cite{Er78,Kr95,Ca95} (as a result of 
the earth's orbit eccentricity \cite{Po69}). The combination of these two 
effects, namely time-dependent spectral distortions, could also be 
observable \cite{Gl87,Kr95}.

	The four pioneering solar neutrino experiments have not observed  
such effects. The three radiochemical experiments  \cite{Da94,GALL,SAGE} 
cannot observe {\em a priori\/} spectral distortions since they detect 
only energy-integrated signals \cite{Ba69}, and do not show  evidence for 
periodic variations of the detected rates associable to just-so oscillations 
(see, e.g., \cite{Pe96,Ca95}).  The neutrino-electron scattering experiment 
Kamiokande \cite{Kami} shows no evidence for distortions in the spectrum 
of the scattered electrons \cite{Kspe} either. However, these are 
low-statistics experiments, and possible vacuum oscillation effects could 
be hidden by the relatively large uncertainties.

	Much higher statistics (thousands of events per year) will be 
collected with the second-generation experiments SuperKamiokande 
\cite{SKam,SKAw} and  Sudbury Neutrino Observatory (SNO) \cite{Sudb,SNOw}. 
These real-time experiments can test variations in the time structure of 
the signal, as well as deviations of the solar neutrino energy spectrum 
from its standard shape. In particular,  information about the $^8$B solar 
neutrino spectrum $\lambda(E)$ \cite{BaLi} can be gained through the 
observation of the electron spectrum \cite{Ba89} from the reactions
\begin{eqnarray}
\nu + e^- &\;\rightarrow\;& \nu' + e^-   \text{ (SuperKamiokande)},
\label{eq:SKreac}\\[2mm]
\nu_e + d &\;\rightarrow\;&  p+p + e^-   \text{ (SNO)}\ .
\label{eq:SNOreac}
\end{eqnarray}

	In this work we study  a specific signal of just-so oscillations 
which involves both time variations and shape distortions of the electron 
kinetic energy spectrum expected at SuperKamiokande and SNO. We propose to
separate the events detected when the earth is nearest to the sun (perihelion
$\pm$~3~months) from those detected  when the earth is farthest from the
sun (aphelion $\pm$~3~months). In Sec.~II we introduce a near-far asymmetry
$A_{NF}$, which is non-zero if just-so oscillations occur. In Sec.~III and 
IV we apply our calculations of $A_{NF}$ to the kinetic energy spectra of
electrons in the SuperKamiokande and SNO experiments respectively. 
We show how the sensitivity to the neutrino oscillation parameters can be 
increased by probing the near-far asymmetry in selected parts of the 
electron energy spectra. In Sec.~V we summarize our work and draw our 
conclusions. Some technical aspects of our calculations are discussed in the 
Appendices A and B.

\section{Near-far asymmetry $A_{NF}$~: General}

	In this section we define the near-far asymmetry $A_{NF}$, 
introduce briefly the notation  for two-family and three-family neutrino
oscillations, and express $A_{NF}$ as a function of the neutrino mass-mixing
parameters. Details of the calculation are reported in Appendix~A.

\subsection{Definition of $A_{NF}$}

	The earth orbit radius, $\ell$, varies periodically around its
average value, $L=1.496\times 10^8$~km, according to
\begin{equation}
\ell=L(1-\varepsilon\cos\vartheta) + {\cal O}(\varepsilon^2)\ ,
\label{eq:ell}
\end{equation}
where $\varepsilon=0.0167$ is the orbit eccentricity, and
\begin{equation}
\vartheta\equiv\frac{2\pi t}{T}
\label{eq:theta}
\end{equation}
is the orbital phase ($T=1\text{ yr}$, $t=0$ at the perihelion).
Variations of $\ell$ with the neutrino production point in the sun are
negligible for our purposes (see Appendix~B).

	Along the earth's orbit, the intercepted neutrino flux $\Phi$ 
varies as:
\begin{equation}
\Phi\propto \frac{L^2}{\ell^2}=1+2\varepsilon\cos\vartheta + 
{\cal O}(\varepsilon^2)\ .
\label{eq:squarelaw}
\end{equation}

	Let us divide the orbit in two parts, a ``near'' half (centered at 
the perihelion, january 4th) and a ``far'' half (centered at the aphelion, 
july 4th):
\begin{mathletters}
\begin{eqnarray}
\text{``near'' semi-orbit} 
&\;=\;& \text{perihelion $\pm$ 3 months}\nonumber\\
&\;\Rightarrow\;&\vartheta\in [-\pi/2,\,\pi/2]\ ,\\[2mm]
\text{``far'' semi-orbit}
&\;=\;&\text{aphelion $\pm$ 3 months}\nonumber\\ 
&\;\Rightarrow\;&\vartheta\in [\pi/2,\,3\pi/2]\ .
\end{eqnarray}
\label{eq:semiorbit}
\end{mathletters}

	If no oscillations occur, the flux of solar $\nu_e$ is subject only
to a geometrical variation [Eq.~(\ref{eq:squarelaw})]. In this case, the 
integrated neutrino rate in the near  semi-orbit is enhanced by a geometric 
factor  $1+4\varepsilon/\pi$:
\begin{equation}
\frac{1}{\pi}
\int_{-\pi/2}^{\pi/2}\!\! d\vartheta \frac{L^2}{\ell^2}=
1+\frac{4\varepsilon}{\pi}+ {\cal O}(\varepsilon^2) \ .
\label{eq:geofactor}
\end{equation}
Analogously, the integrated neutrino rate in the far  semi-orbit is 
suppressed by a geometric factor $1-4\varepsilon/\pi$.

	It is useful to factorize out this overall geometric correction 
from the observations and define%
\footnote{	In principle, a real-time signal could be  instantaneously 
corrected for the geometric factor $L^2/\ell^2$, if also the background 
were known in real time and subtracted.  However, background subtraction 
is better defined for  time-integrated signals. Therefore, we prefer to 
factorize the half-year averaged geometric correction $1\pm4\epsilon/\pi$ 
from the integrated rate, instead of factorizing the instantaneous 
correction $L^2/\ell^2$ from the real-time rate.} 
\begin{mathletters}
\begin{eqnarray}
N&=&\frac{\text{No.~of events in the near semi-orbit}}{1+4\varepsilon/\pi}
\label{eq:N}\ ,\\[2mm]
F&=&\frac{\text{No.~of events in the far semi-orbit}}{1-4\varepsilon/\pi}\ .
\label{eq:Far}
\end{eqnarray}
\label{eq:counts}
\end{mathletters}

	``Event'' is referred here to as the observation of a solar neutrino 
induced electron. The relevant reactions are Eq.~(\ref{eq:SKreac}) for 
SuperKamiokande  (neutrino scattering) and Eq.~(\ref{eq:SNOreac}) 
for SNO (neutrino absorption). Both experiments can measure the kinetic
energy, $T$, of the scattered or  recoil electrons through the Cherenkov
light. The ``No.~of~events'' in Eq.~(\ref{eq:counts}) may refer either to 
the whole electron energy spectrum, $s(T)$, or to a selected part of it. 
The first option has the advantage of higher statistics; the second is 
useful to study possible time-dependent spectral distortions. The formalism 
that will be used in  Sections II~B and II~C  is the same in both cases. 
Further specifications about the electron  energy spectrum will be made in 
Sections~III and IV.

	We define a near-far asymmetry as 
\begin{equation}
A_{NF}=\frac{N-F}{N+F}\ .
\label{eq:Asy}
\end{equation}
The statistical error of $A_{NF}$, as derived by applying the binomial
statistics to the ``near'' and ``far'' classes of events, is 
$1/\sqrt{N_{\rm tot}}$,  where $N_{\rm tot}=N+F$. Sistematic errors can be 
expected to cancel to a large fraction in a ratio of rates such as 
$(N-F)/(N+F)$.

	The asymmetry $A_{NF}$ is zero if no neutrino oscillations occur, 
or if oscillations independent of $\ell$ take place. Therefore, a non-zero 
value of $A_{NF}$ would be a signature of just-so oscillations. Notice 
that $A_{NF}$ is independent of the solar model and, in particular, of 
the absolute $^8$B neutrino flux.

\subsection{Vacuum oscillations: Notation}

	In the presence of just-so oscillations, the asymmetry $A_{NF}$ is 
a function of the neutrino mass-mixing parameters. We consider two cases: 
two-family $(2\nu)$ oscillations, and three-family $(3\nu)$ oscillations 
with only one relevant mass scale.

	In the two-family case, the electron neutrino $\nu_e$ is a mixture of
two mass eigenstates $\nu_1$ and $\nu_2$, with masses $m_1$ and $m_2$
respectively:
\begin{eqnarray}
\nu_e = c_\omega\nu_1 + s_\omega \nu_2\ ,\\[2mm]
\delta m^2 =  |m^2_2-m^2_1|\ ,
\end{eqnarray}
where $\omega$ is the $\theta_{12}$ mixing angle, 
$s_\omega=\sin\omega$, and $c_\omega=\cos\omega$.

	The $\nu_e$ survival probability, 
$P^{2\nu}=P^{2\nu}(\nu_e\rightarrow\nu_e)$, is then given by
\begin{equation}
P^{2\nu}=1-2s^2_\omega c^2_\omega + 2s^2_\omega c^2_\omega\cos k\ell\ ,
\label{eq:P2nu}
\end{equation}
where
\begin{eqnarray}
k=\frac{2\pi}{\lambda_\nu} &=& \text{ wave number}\ ,\\[2mm]
\lambda_\nu=\frac{4\pi E}{\delta m^2}&=& \text{ oscillation length}\ ,
\end{eqnarray}
and $E$ is the neutrino energy.

	In the case of three-family neutrino oscillations, the state $\nu_e$ 
is a mixture of $\nu_1$, $\nu_2$, and $\nu_3$:
\begin{equation}
\nu_e=c_\phi c_\omega \nu_1 + c_\phi s_\omega \nu_2 + s_\phi \nu_3\ ,
\label{eq:P3nu}
\end{equation}
where $\omega=\theta_{12}$ and $\phi=\theta_{13}$ in the standard 
parametrization of the mixing matrix \cite{Ku87,PDGR}.

	We make the simplificative hypothesis that the mass gap between
$\nu_3$ and the doublet $(\nu_1,\,\nu_2)$ is very large:
$m^2\equiv |m^2_3-m^2_1| \gg |m^2_2-m^2_1|\equiv \delta m^2$. 
This hypothesis holds, for instance, if one assumes that $m^2$ is in the 
range ($\gtrsim 10^{-3}$ eV$^2$) probed by accelerator, reactor, and
atmospheric neutrino experiments \cite{Larg}. With this assumption the 
$m^2$-driven fast oscillations are averaged away, and the $3\nu$ survival 
probability is given by (see, e.g., \cite{Larg})
\begin{equation}
P^{3\nu}=c^4_\phi P^{2\nu} + s^4_\phi\ ,
\end{equation}
with $P^{2\nu}$ as in Eq.~(\ref{eq:P2nu}).

	A useful representation of the $3\nu$ parameter space 
$(\delta m^2,\,\omega,\,\phi)$ has been given in \cite{Tria}
in terms of a ``unitarity triangle.'' We refer the reader to \cite{Tria} 
for a description of this representation, which will be used to display 
some of the results in Sections~III and IV.

	In summary:
\begin{mathletters}
\begin{eqnarray}
\text{no oscillations}    &\;\Rightarrow\;&
A_{NF}=0\ ,\\[2mm]
\text{$2\nu$ oscillations}&\;\Rightarrow\;&
A^{2\nu}_{NF}=A^{2\nu}_{NF}(\delta m^2,\,\omega)\ ,\label{eq:A2}\\[2mm]
\text{$3\nu$ oscillations}&\;\Rightarrow\;&
A^{3\nu}_{NF}=A^{3\nu}_{NF}(\delta m^2,\,\omega,\,\phi)\ .\label{eq:A3}
\end{eqnarray}
\end{mathletters}

	In the next Section we give explicit formulas for  
Eqs.~(\ref{eq:A2}) and (\ref{eq:A3}).

\subsection{Calculation of $A_{NF}$}

	The rates $N$ and $F$ in Eq.~(\ref{eq:counts}) are integrated over 
a half-year and over all, or part, of the electron energy spectrum. It 
follows that the calculation of $A_{NF}$ involves a time integration
over the near and far semiorbits, as well as an energy  integration 
weighted by the $^8$B neutrino spectrum, $\lambda(E)$, and by the 
$\nu_e$ and $\nu_x$ ($x\neq e$) cross sections, $\sigma_e(E)$ and 
$\sigma_x(E)$ respectively.

	In the calculation, we discard terms of ${\cal O}(\varepsilon^2)$, 
but we keep all powers in $\varepsilon k L$ since $k L$ may be large.
We also neglect the very small smearing effect over the neutrino 
production region (see Appendix~B).

	It turns out that the time integration in the two semiorbits can 
be done analytically, leaving only a numerical integration over $E$ to be 
performed. We refer the reader to Appendix~A for the derivation of the 
final results, that can be expressed in compact form as
\begin{equation}
A^{2\nu}_{NF}=\frac{2s^2_\omega c^2_\omega G(\delta m^2)}
{1-2s^2_\omega c^2_\omega F(\delta m^2)}\ ,
\label{eq:A2nu}
\end{equation}
for $2\nu$ oscillations, and as
\begin{equation}
A^{3\nu}_{NF}=\frac{2c^4_\phi s^2_\omega c^2_\omega G(\delta m^2)}
{1-2 s^2_\phi c^2_\phi R - 2c^4_\phi s^2_\omega c^2_\omega F(\delta m^2)}\ .
\label{eq:A3nu}
\end{equation}
for $3\nu$ oscillations.

The functions $F(\delta m^2)$ and $G(\delta m^2)$ and the factor $R$
in  Eqs.~(\ref{eq:A2nu}) and (\ref{eq:A3nu}) are defined as
\begin{eqnarray}
F(\delta m^2) &=& \frac
{\int\!dE\,\lambda(E)\,[\sigma_e(E)-\sigma_x(E)]\, f(\delta m^2/E)}
{\int\!dE\,\lambda(E)\,\sigma_e(E)}\ ,\label{eq:F}\\[2mm]
G(\delta m^2) &=& \frac
{\int\!dE\,\lambda(E)\,[\sigma_e(E)-\sigma_x(E)]\, g(\delta m^2/E)}
{\int\!dE\,\lambda(E)\,\sigma_e(E)}\ ,\label{eq:G}\\[2mm]
R &=& \frac
{\int\!dE\,\lambda(E)\,[\sigma_e(E)-\sigma_x(E)]}
{\int\!dE\,\lambda(E)\,\sigma_e(E)}\ ,\label{eq:R}
\end{eqnarray}
where $f$ and $g$ are  non-dimensional functions of $\delta m^2/E$ ($=2k$):
\begin{eqnarray}
f &=& 1-c J_0(\varepsilon k L)
 -2\varepsilon s J_1(\varepsilon kL)
+\frac{4\varepsilon}{\pi} s {\bf H}_0 (\varepsilon kL)
\ ,\label{eq:f}\\[2mm]
g &=&\frac{4\varepsilon}{\pi}c
+ s {\bf H}_0(\varepsilon kL)
 -2\varepsilon c {\bf H}_1(\varepsilon kL)
-\frac{4\varepsilon}{\pi} c J_0 (\varepsilon kL)
\label{eq:g}\ ,
\end{eqnarray}
with $s=\sin kL$, $c=\cos kL$. In Eqs.~(\ref{eq:f}) and  (\ref{eq:g}),  
$J_{0,1}$ and ${\bf H}_{0,1}$ are the Bessel and Struve functions 
\cite{Grad,Abra}  of order 0 and 1, respectively. Computer routines 
for the numerical calculation of these special functions can be found 
in \cite{CERN}.

	Figure~1 shows the functions $f$ and $g$ in the relevant range
$\delta m^2/E\in[10^{-12},\,10^{-8}]$  eV$^2$/MeV. These functions are 
detector-independent. The calculation of $F$ and $G$ from  Eqs.~(\ref{eq:F}) 
and (\ref{eq:G}) requires instead detector-dependent ingredients such as 
the interaction cross-sections. These ingredients are discussed in 
Sec.~III and IV for  SuperKamiokande and SNO, respectively.

\section{$A_{NF}$ in SuperKamiokande}

	In this section we calculate the expected values of $A_{NF}$ for 
SuperKamiokande in the presence of $2\nu$ and $3\nu$ oscillations.

\subsection{Detector parameters}

	The SuperKamiokande experiment \cite{SKam} makes use of a 22 kton 
(fiducial volume) water-Cherenkov detector to observe the electrons 
scattered by solar  neutrinos through reaction~(\ref{eq:SKreac}).

	The {\em measured\/} kinetic energy of the electron, $T$, is 
expected to be distributed around the {\em true\/} electron energy, $T'$, 
according to the following energy resolution function \cite{Ba89}:
\begin{equation}
R(T,\,T')=\frac{1}{\Delta_{T'}\sqrt{2\pi}}
\exp\left[-\frac{(T-T')^2}{2\Delta_{T'}^2}\right]\ ,
\label{eq:resol}
\end{equation}
where the energy-dependent one-sigma width $\Delta_{T'}$ scales as 
$\sqrt{T'}$ due to the photon statistics,
\begin{equation}
\Delta_{T'}=\Delta_{10}\sqrt{\frac{T'}{\text{10 MeV}}}\ ,
\end{equation}
and the nominal width at 10 MeV is $\Delta_{10}=1.6$ MeV  \cite{SKam}.

	If the experiment works as expected, a high signal-to-noise ratio
will be achieved for $T\gtrsim 5$ MeV \cite{Suzu}. We adopt 
$T_{\rm min}=5$ MeV as a nominal threshold in our analysis.

\subsection{Neutrino cross-section and electron spectrum}

	The neutrino cross section for the reaction~(\ref{eq:SKreac}) is 
known to the first order in the radiative corrections \cite{Sirl}, both for 
$\nu=\nu_e$ and for $\nu=\nu_x$ $(x\neq e)$.

	The standard electron spectrum $s(T)$ expected in SuperKamiokande
in the absence of oscillations is given by
\begin{equation}
s(T)=\int \!\!dE\,\lambda(E) \int_0^{T'_{\rm max}}\!\! dT'\,
R(T,\,T')\, 
\frac{d\sigma_e(E,\,T')}{dT'}\ ,
\end{equation}
where $\lambda(E)$ is the neutrino spectrum, $T'_{\rm max}$ is the maximum 
{\em true\/} kinetic energy allowed by kinematics  \cite{Ba89}, $R(T,\,T')$ 
is given in Eq.~(\ref{eq:resol}), and $d\sigma_e/dT'$ is the $\nu_e$ 
differential cross section  \cite{Ba89,Sirl}.

	Figure~2 shows the standard spectrum $s(T)$ for SuperKamiokande.
As we will see in Sec.~III~C, it is also useful to consider selected  parts 
of the electron energy spectrum (bins).  An illustrative choice of bins is
shown in Fig.~2. The first five bins have a width of 1 MeV.  The sixth bin 
collects all events with $T\ge 10$ MeV. In the presence of oscillations, 
we will calculate the asymmetry $A_{NF}$ using both the total number of 
events in the  spectrum, and the  number of events collected in each of the
six bins shown in Fig.~2. The loss of statistics is then traded for an 
increased sensitivity to the neutrino mass-mixing parameters (see Sec.~III~C).

	It is useful to introduce a ``reduced'' $\nu_e$ cross-section for
the $i$-th bin, $\sigma_{e,i}(E)$, defined as the cross section for a 
$\nu_e$ of energy $E$ to produce an electron with a  {\em measured\/} 
energy falling in the $i$-th bin range  $[T_{i,\rm min},\,T_{i,\rm max}]$:
\begin{equation}
\sigma_{e,i}(E)=\int_{T_{i,\rm min}}^{T_{i,\rm max}}\!\!dT
\int_0^{T'_{\rm max}}\!\! dT'\, R(T,\,T')\, \frac{d\sigma(E,\,T')}{dT'}
\label{eq:reduce}
\end{equation}
An analogous definition can be given for the $\nu_x$ reduced
cross-sections, $\sigma_{x,i}(E)$.

	The total cross section $\sigma_e$ for producing an electron with 
measured energy above the nominal threshold ($T_{\rm min}=5$~MeV) and
including the finite energy resolution is:
\begin{equation}
\sigma_{e}(E)=\sum_{i=1}^6 \sigma_{e,i}(E)\ . 
\end{equation}

	The total cross sections  $\sigma_{e}(E)$ and  $\sigma_{x}(E)$, or 
the binned cross sections  $\sigma_{e,i}(E)$ and  $\sigma_{x,i}(E)$, enter 
the calculation of $A_{NF}$ through the integrals in 
Eqs.(\ref{eq:F})--(\ref{eq:R}), where they are always weighted by
the $^8$B neutrino spectrum $\lambda(E)$.

	In Fig.~3 we show the products  $\lambda(E)\sigma_{e}(E)$ (upper 
panel) and  $\lambda(E)\sigma_{e,i}(E)$ (lower panel) as a function of $E$. 
The curve in the upper panel thus represents the $\nu_e$ spectrum 
contribution to the whole electron spectrum in SuperKamiokande. 
Analogously, the  curves in the lower panel represent the  $\nu_e$ spectrum 
contribution to each of the six bins shown in Fig.~2.%
\footnote{	Curves of $\lambda\sigma_x$ (not shown in Fig.~3) would be 
		similar to the $\lambda\sigma_e$ curves, but scaled
		by a factor $\sim 1/6$ in height.}
Notice that, although the width of the first five electron energy bins
is 1 MeV, the corresponding width of the $\lambda(E)\sigma_{e,i}(E)$
curves is much larger, indicating that there is no tight relation between 
the measured electron energy $T$ and the energy $E$ of the parent neutrino. 
This is due both to  the smearing effect of the energy resolution 
function [Eq.~(\ref{eq:resol})] and to the fact that the differential
cross section $d\sigma/dT$ for $\nu$-$e$ scattering is rather flat
in $T$ \cite{Ba89}. The finite  energy resolution  also allows non-zero
values of $\lambda(E)\sigma_{e,i}(E)$ for $E<T_{i,\rm min}$.
Nevertheless, from Fig.~3 we learn that it is possible  to ``tune,'' to 
some extent, the typical neutrino energy relevant to $A_{NF}$ by choosing 
the appropriate bin in the electron energy spectrum.

\subsection{$A_{NF}$ with $2\nu$ oscillations}

	Figure~4 shows curves of iso-$A_{NF}$ ($\times 100$) for the 
unbinned (whole spectrum) case, in the presence of $2\nu$ oscillations. 
The curves are displayed in the usual $(\sin^2 2\omega,\,\delta m^2)$ plane, 
and are separated by steps of 0.5\% in $A_{NF}$.

	The maximum absolute value of $A_{NF}$ ($\sim 2.5\%$) is reached 
for $\sin^2 2\omega\simeq 1$ and $\delta m^2\simeq 10^{-10}$ eV$^2$. If the 
statistical errors dominate, a near-far effect of $2\%$ could be detected at 
SuperKamiokande after the first  year of operation.

	The near-far asymmetry can be either positive or negative 
(solid and dotted curves, respectively).  The positive (negative) values of 
$A_{NF}$ correspond to an energy-averaged $\nu_e$ survival probability 
$\langle P\rangle$ decreasing (increasing) with the sun-earth distance. 
$A_{NF}$ is $\sim 0$ at  the stationary points of $\langle P\rangle$, 
corresponding approximately to
\begin{equation}
	L\simeq n\,\frac{\langle \lambda_\nu \rangle }{2}\ ,
\label{eq:zeros}
\end{equation}
where $\langle \lambda_\nu \rangle = 4\pi \langle E \rangle/\delta m^2$ is 
the oscillation length in terms of the typical neutrino energy 
$\langle E \rangle$ ($\sim 10$ MeV for Fig.~4) and $n$ is an integer.
The near-far asymmetry tends to zero also for large $\delta m^2$
(fast oscillations), for very small $\delta m^2$ (oscillation lengths 
much longer than $L$), and for small $\sin^2 2\omega$ (small neutrino mixing).

	If Eq.~(\ref{eq:zeros}) is fulfilled for a certain value of 
$\delta m^2$, no near-far asymmetry can be observed, independently of the 
value of $\sin^2 2\omega$. However, as noticed in the previous section,
one can change the value of $\langle E\rangle$ (and thus the position
of the zeros of $A_{NF}$) by selecting the events in a specific bin of 
the electron energy spectrum.

	In Fig.~5 we show curves of iso-$A_{NF}$ calculated for the six bins
indicated in Fig.~2. The typical neutrino energy increases from bin 1 to 
bin 6 (see Fig.~3) and correspondingly the sensitivity to higher $\delta m^2$ 
increases. The positions of the zeros of $A_{NF}$ are also shifted from bin 
to bin, so that by using two or more bins no ``holes'' in the 
$\delta m^2$-sensitivity are left. Notice that  the near-far asymmetry can 
reach values as high as 3.5\% in favorable cases of large mixing, and that 
the sensitivity to small mixing is increased in some bins.

	The comparison of Figs.~4 and 5 shows that selecting some intervals
(bins) of the electron energy spectrum helps to reach higher sensitivities
to the neutrino oscillation parameters. Of course, the statistical  (and 
possibly the systematic) uncertainties also increase when a subset of the 
total event sample is used. An optimal sensitivity-to-uncertainty ratio can 
be assessed only after the detector performances and the results
of the SuperKamiokande experiment are known in detail.

\subsection{$A_{NF}$ with $3\nu$ oscillations}

	As discussed in Sec.~II~B, we study $3\nu$ oscillations in
the simple case of one relevant mass scale. The corresponding parameter 
space $(\delta m^2,\,\omega,\,\phi)$ can be represented, for any fixed 
value of $\delta m^2$, through the ``triangular'' representation 
introduced in \cite{Tria}.

	Figure~6 shows curves of iso-$A_{NF}$\ $(\times 100)$ in the 
triangular representation for $\delta m^2=10^{-10}$ eV$^2$ and 
$\delta m^2=5.5\times 10^{-11}$. The calculations refer to the unbinned 
case (whole spectrum).  We recall \cite{Tria} that  a generic point in 
the triangle represents the electron-neutrino state $\nu_e$, while the 
upper, lower right, and lower left corners represent the mass eigenstates 
$\nu_3$, $\nu_2$, and $\nu_1$ respectively.  The triangle is mapped by 
the (non-orthogonal) coordinates $\sin^2\omega$ and $\sin^2\phi$. The 
asymmetry $A_{NF}$  is maximal (in absolute value) on the lower side of 
the triangle, corresponding to pure $2\nu$ oscillations between $\nu_1$ 
and $\nu_2$.  For $\phi>0$, the $\nu_e$-$\nu_3$ mixing tends to suppress 
the near-far effect, since by hypothesis the oscillations driven by 
massive state $\nu_3$ are so fast  to be averaged out, giving a 
time-independent contribution. In short, in order to have a large 
near-far asymmetry for $3\nu$ oscillations, the $\omega$-mixing must 
be maximal and the  $\phi$-mixing must be relatively small.

	In Fig.~6 we have shown only two representative values of 
$\delta m^2$. The range of $\delta m^2$ to which $A_{NF}$ is sensitive
for $3\nu$ oscillations would be practically the same as for $2\nu$ 
oscillations. The zeros of  $A^{3\nu}_{NF}$ and $A^{2\nu}_{NF}$ are 
also reached at  the same values of $\delta m^2$ [compare 
Eqs.~(\ref{eq:A2nu}) and (\ref{eq:A3nu})]. These ``holes'' in the 
$\delta m^2$-sensitivity can be ``closed,'' as in the $2\nu$ case,
by repeating the asymmetry measurement in selected bins of the electron
energy spectrum.

\section{$A_{NF}$ in SNO}

In this section we calculate the expected values of $A_{NF}$ at SNO in the 
presence of $2\nu$ and $3\nu$  oscillations.

\subsection{Detector parameters}

	The SNO experiment \cite{Sudb} makes use of a 1 kton  heavy-water 
Cherenkov detector to observe the recoil electrons  from  neutrino absorption
in deuterium [Eq.~(\ref{eq:SNOreac})].

	The discussion of the detector parameters would be very similar
to SuperKamiokande (Sec.~III~A) and is not repeated here. We only report 
the adopted (expected) values for the resolution width at 10~MeV, 
$\Delta_{10}=1.1$ MeV, and for the energy threshold, $T_{\rm min}=5$ MeV 
\cite{Li96}.

\subsection{Neutrino cross section and electron spectrum}

	The cross section for neutrino absorption in deuterium has been 
studied in detail in several papers (see \cite{Ku94} and references therein). 
No relevant uncertainties are recognized at solar neutrino energies 
\cite{Ku94,Li96}, apart from a $\sim \pm 5\%$ error in the overall 
normalization  that is irrelevant in a ratio of rates like $A_{NF}$. A 
computer code for the calculation of the $\nu$-$d$ differential
cross section is available in \cite{wwwB}.

	The differential cross section  $d\sigma_e/dT$ for 
reaction~(\ref{eq:SNOreac})  is sharply peaked in the electron energy 
(see, e.g., \cite{Li96}). This feature represents a considerable advantage 
with respect to $\nu$-$e$ scattering in SuperKamiokande 
[reaction~(\ref{eq:SKreac})], since it results in a tighter correlation 
between the electron energy $T$  and the parent neutrino energy $E$.

	Figure~7  shows the standard (i.e., no oscillation) electron 
spectrum for SNO. In Fig.~7 we also show, as in Fig.~3, the energy bins 
chosen to illustrate to near-far asymmetry effect in selected
parts of the spectrum.

	Figure~8 illustrates the  $T$-$E$ correlation in SNO. This figure 
is analogous to Fig.~4, but with the appropriate SNO neutrino 
cross-sections (defined in analogy to Eq.~(\ref{eq:reduce})). 
By changing the electron energy bin, the  spectrum $\lambda(E)\sigma_e(E)$
of parents neutrinos contributing to that bin changes considerably. 
The corresponding variations in SuperKamiokande (Fig.~4) are less 
pronounced. Therefore we expect a greater sensitivity
to near-far effects in SNO.

	Another advantage of SNO is that the reaction~(\ref{eq:SNOreac}) can
be initiated only by electron neutrinos: $\sigma_x=0$ $(x\neq e)$. 
It follows that the factor $R$ defined in Eq.~(\ref{eq:R}) is equal to 1.

\subsection{$A_{NF}$ with $2\nu$ oscillations}

	Figure~9 shows curves of iso-$A_{NF}$ for SNO (unbinned case),
in the $2\nu$ oscillation plane $(\sin^2 2\omega,\,\delta m^2)$. The 
contours are separated by steps of $1\%$ in $A_{NF}$.

	The shape of the curves in Fig.~9 is similar to Fig.~4, since the
typical neutrino energy is $\langle E\rangle\sim 10$ MeV as in  
SuperKamiokande. However, the absolute value of the asymmetry in SNO is
about twice as large as in SuperKamiokande, because of the tighter
correlation between the observed electron spectrum and the parent neutrino
spectrum, as discussed in Sec.~IV~B. The maximum absolute value of $A_{NF}$ 
is $\sim 6.3\%$.

	Figure~10 shows curves of iso-$A_{NF}$ calculated for the six 
representative bins shown in Fig.~7. As for SuperKamiokande, we note that 
by changing bin the $\delta m^2$-position of the $A_{NF}$ zeros  changes, 
and higher sensitivities to low mixing angles can also be achieved.

	A comparison of Figs.~4, 5, 9, and 10, shows that the SuperKamiokande 
and the SNO experiment can probe, through measurements of the near-far 
asymmetry, the range  
$2.5 \times 10^{-11} \lesssim \delta m^2 \lesssim 9 \times 10^{-10}$
eV$^2$ for sufficiently large mixing.

\subsection{$A_{NF}$ with $3\nu$ oscillations}

	Figure~11 shows curves of iso-$A_{NF}$ for SNO, in the presence
of $3\nu$ oscillations, for two representative values of $\delta m^2$.

	In Fig.~11  the same triangular representation of Fig.~6 is used. 
The shape of the contours in Fig.~11 (SNO) and Fig.~6 (SuperKamiokande) 
are similar, but the near-far asymmetry is larger in the SNO experiment, 
for the same reason as in the $2\nu$ case (see Sec.~IV~C).

\section{Summary and conclusions}

	We have studied  the signals of just-so oscillations that can be 
observed  in the SuperKamiokande and SNO solar neutrino experiments by 
separating the events detected when the earth is nearest to the sun 
(perihelion $\pm$~3~months) from those detected when the earth is farthest 
from the sun (aphelion $\pm$~3~months).

	We have calculated the asymmetry $A_{NF}$ between the near and far 
signals by factorizing out the trivial geometrical variation of the signal,
and using the entire electron energy spectrum as well as representative
spectrum bins. The calculations involve the integration of the  $\nu_e$ 
survival probability over time and energy. The time integration over 
half-year can be performed analitically. Compact expressions for the near-far
asymmetry have been given for $2\nu$ and $3\nu$ oscillations. The value of 
$A_{NF}$ is solar model independent, and is different from zero if just-so 
oscillations occur.

 	In the case of $2\nu$ oscillations, it has been shown that 
measurements of the near-far asymmetry at SNO and SuperKamiokande can probe 
the range  $2.5 \times 10^{-11} \lesssim \delta m^2 \lesssim 9\times 10^{-10}$
eV$^2$ for sufficiently large values of the mixing angle $\omega=\theta_{12}$. 
A similar range of $\delta m^2$ and $\omega$ can be probed in the presence 
of $3\nu$ oscillations with one relevant mass scale, provided that the 
second mixing angle $\phi=\theta_{13}$  is not large.

	The SNO experiment appears to be about twice as sensitive  as 
SuperKamiokande to the near-far asymmetry, due to the different energetics of 
the $\nu$-$d$ absorption and $\nu$-$e$ scattering reactions.

	In both experiments, measurements of $A_{NF}$ in selected bins of 
the electron energy spectrum may increase the sensitivity to the neutrino
oscillation parameters. Since the selection of a bin implies a loss of 
statistics, the net gain of binning the spectrum in measurements of $A_{NF}$  
can  be assessed only after the experiments have run for some time and  
the data and their uncertainties are well understood.

\acknowledgments

	We thank P.~I.~Krastev for useful discussions. One of us (B.F.) 
thanks the Dipartimento di Fisica and Sezione INFN di Bari for kind 
hospitality.  The work of E.L.\ was supported in part by  INFN and in 
part by the Institute for Advanced Study through a Hansmann membership.  
This work has been performed under the auspices of the European 
Theoretical Astroparticle Network (TAN).

\appendix
\section{Time and energy integration in $A_{NF}$}

	In this Appendix we prove the basic 
Eqs.~(\ref{eq:A2nu}) and (\ref{eq:A3nu}) for the near-far asymmetry.

	In the presence of oscillations, the expected value of the 
quantity $N$ in Eq.~(\ref{eq:N}) is given by the sum of the $\nu_e$ and 
$\nu_x$ contributions
\begin{eqnarray}
N&=&\int\!dE\,\lambda(E)\sigma_e(E)\overline{P}_N(E)\nonumber\\
& & \mbox{}+\int\!dE\,\lambda(E)\sigma_x(E)(1-\overline{P}_N(E)),
\label{eq:Near}
\end{eqnarray}
where $E$ is the neutrino energy, $\lambda(E)$ is the $^8$B neutrino 
spectrum, $\sigma_e$ and $\sigma_x$ are the $\nu_e$ and $\nu_x$ ($x\neq e$) 
interaction cross-sections,%
\footnote{	It is understood that the cross sections are already corrected
		for threshold and  resolution effects in the electron 
		spectrum. The cross sections may refer either to the entire 
		electron spectrum (above threshold) or to a specific bin. 
		See Secs.~III~B and IV~B for details.}
and $\overline{P}_N$ is the $\nu_e$ survival probability $P(E,\,\vartheta)$
averaged over the near semiorbit:
\begin{equation}
\overline{P}_N(E)=\frac
{\displaystyle\frac{1}{\pi}\int_{-\pi/2}^{\pi/2}\!\!d\vartheta\,
\frac{L^2}{\ell^2}\,P(E,\,\vartheta)}
{\displaystyle\frac{1}{\pi}\int_{-\pi/2}^{\pi/2}\!\!d\vartheta\,
\frac{L^2}{\ell^2}}\ .
\label{eq:PN}
\end{equation}

	In the above equation, $L^2/\ell^2$ is given by 
Eq.~(\ref{eq:squarelaw}), and the denominator on the right hand side is 
equal to  \mbox{$1+4\varepsilon/\pi$} [Eq.~(\ref{eq:geofactor})]. For 
$P=P^{2\nu}$  or $P=P^{3\nu}$ [Eqs.~(\ref{eq:P2nu}) and (\ref{eq:P3nu})]
the numerator can be calculated analitically with the help of the following 
integrals \cite{Grad,Abra}:
\begin{eqnarray}
\frac{2}{\pi}\int_0^{\pi/2}\!\!\!d\vartheta
\cos(x\cos\vartheta)&=&J_0(x)\ ,\\[2mm]
\frac{2}{\pi}\int_0^{\pi/2}\!\!\!d\vartheta
\sin(x\cos\vartheta)&=&{\bf H}_0(x)\ ,\\[2mm]
\frac{2}{\pi}\int_0^{\pi/2}\!\!\!d\vartheta
\cos\vartheta\sin(x\cos\vartheta)&=&J_1(x)\ ,\\[2mm]
\frac{2}{\pi}\int_0^{\pi/2}\!\!\!d\vartheta
\cos\vartheta\cos(x\cos\vartheta)&=&\frac{2}{\pi}-{\bf H}_1(x)\ ,
\end{eqnarray}
where $J_n$ and ${\bf H}_n$ are the Bessel and Struve functions 
\cite{Grad,Abra} of order $n$:
\begin{eqnarray}
J_n(x)&=&\left(\frac{1}{2}x\right)^n\sum_{m=0}^{\infty}
\frac{(-1)^m\left(\frac{1}{2}x\right)^{2m}}{\Gamma(m+1)\Gamma(n+m+1)}\ ,\\[2mm]
{\bf H}_n(x)&=&\left(\frac{1}{2}x\right)^{n+1}\sum_{m=0}^{\infty}
\frac{(-1)^m\left(\frac{1}{2}x\right)^{2m}}
{\Gamma(m+\frac{3}{2})\Gamma(n+m+\frac{3}{2})}\ .
\end{eqnarray}

In particular, in the $2\nu$ oscillation case the result of the integration 
is:%
\footnote{	Terms of ${\cal O}(\varepsilon^2)$ or higher are neglected.
		However, all powers of $\varepsilon k \ell$ are kept, since
		$\varepsilon k \ell$ can be of ${\cal O}(1)$.} 

\begin{eqnarray}
\frac{1}{\pi}\int^{\pi/2}_{-\pi/2}\!\!d\vartheta\,\frac{L^2}{\ell^2}
\lefteqn{P^{2\nu}(E,\vartheta)=}\nonumber\\
& &  1+\frac{4\varepsilon}{\pi}
-2s^2_\omega c^2_\omega (1-cJ_0-2\varepsilon s J_1)\nonumber\\
& & +2s^2_\omega c^2_\omega 
\left[ s{\bf H}_0 -2\varepsilon c {\bf H}_1 + \frac{4\varepsilon}{\pi}
(c-1)\right] ,
\label{eq:bigintegral}
\end{eqnarray}
where $s=\sin kL$, $c=\cos kL$, and the argument of  the functions $J_n$ 
and ${\bf H}_n$ is $\varepsilon k L$. 

	From Eqs.~(\ref{eq:PN}), (\ref{eq:bigintegral}),  and 
(\ref{eq:geofactor}), it follows that 
\begin{equation}
\overline{P}_{N}^{2\nu}=1-2 s^2_\omega c^2_\omega (f-g) + 
{\cal O}(\varepsilon^2)\ ,\label{eq:1st}
\end{equation}
where $f$ and $g$ are defined as in Eqs.~(\ref{eq:f}) and (\ref{eq:g}).
The function $f$ ($g$) is even (odd)  in $\varepsilon$.

	The analogous result for the far semiorbit can be simply
obtained with the replacement $\varepsilon\to-\varepsilon$:
\begin{equation}
\overline{P}_{F}^{2\nu}=1-2 s^2_\omega c^2_\omega (f+g) + 
{\cal O}(\varepsilon^2)\ .
\end{equation}

	The derivation for the $3\nu$ oscillation case  is very similar to 
the $2\nu$  case and gives:
\begin{eqnarray}
\overline{P}_{N}^{3\nu}
&=& c^4_\phi+s^4_\phi-2c^4_\phi s^2_\omega c^2_\omega (f-g)\ ,\\[2mm]
\overline{P}_{N}^{3\nu}
&=& c^4_\phi+s^4_\phi-2c^4_\phi s^2_\omega c^2_\omega (f+g)\ .
\label{eq:4th}
\end{eqnarray}

	Finally, we group the cross-sections and the probabilities in the
expression of near-far asymmetry [see Eqs.~(\ref{eq:Asy}) and 
(\ref{eq:Near})]:
\begin{equation}
A_{NF}=\frac
{\int \!dE\, \lambda (\sigma_e-\sigma_x)(\overline{P}_N-\overline{P}_F)}
{2\int\! dE\, \lambda \sigma_e +\int\! dE\, \lambda (\sigma_e-\sigma_x)
(\overline{P}_N+\overline{P}_F-2)}\ ,
\end{equation}
and  divide the numerator and the denominator of the above fraction
by $\int\!dE\,\sigma_e$. Then, using Eqs.~(\ref{eq:1st})--(\ref{eq:4th}) 
and the definitions in Eqs.~(\ref{eq:F})--(\ref{eq:R}) one easily obtains
Eqs.~(\ref{eq:A2nu}) and (\ref{eq:A3nu}).

\section{Integration over the neutrino production region}

	In the calculation of $A_{NF}$ we have  neglected the effect of 
smearing the $\nu_e$ survival probability $P$ over $^8$B neutrino production 
region in the sun \cite{Pi95}. Already in  \cite{Po67} it was recognized that
this effect is negligible for just-so oscillations, since the typical 
oscillation length is much larger than the  production region (see also 
\cite{Kr96}).  Here we rederive and discuss  this  result in more detail.

	Let us chart the neutrino production region with polar coordinates
$(r,\,\alpha)$, where the angle $\alpha$ is zero along the line joining
the center of the sun ($r\equiv0$) to the detector. (The second polar angle
is integrated out for the cylindrical symmetry of the problem.) The 
distance $\ell'$ between the generic neutrino production point and
the detector is then given by
\begin{equation}
\ell'=\ell -r\cos\alpha\ .
\label{eq:ell1}
\end{equation}

	We find that the $^8$B neutrino source density in the standard 
solar model \cite{Pi95} is approximated very well by the following 
Gaussian distribution:
\begin{equation}
\rho(r)=\frac{1}{(2\pi\sigma^2)^{3/2}}\exp\left(-\frac{1}{2}
\frac{r^2}{\sigma^2}\right)\ ,
\label{eq:rho}
\end{equation}
where $\sigma=0.0313\, R_\odot$ ($R_\odot=6.96\times 10^5$ km).  Since 
$\sigma \ll  R_\odot$, one can effectively take  $r=\infty$ as upper 
limit in the radial integration, instead of $r=R_\odot$.

	The normalization of $\rho(r)$ is fixed by:
\begin{equation}
2\pi\int_{-1}^{+1}\!d\cos\alpha\int_0^{\infty}\!dr\,r^2\rho(r) = 1
\end{equation}

	Given Eq.~(\ref{eq:ell1})  and the functional form (\ref{eq:rho}) 
for $\rho$, the smearing of the neutrino oscillation factor $\cos k\ell $ 
[Eq.~(\ref{eq:P2nu})]  over the production region can be performed 
analytically:
\begin{eqnarray}
\langle\cos k\ell\rangle &\equiv& 
2\pi\int_{-1}^{+1}\!d\cos\alpha\int_0^{\infty}\!dr\,r^2\rho(r) \cos k\ell'
\nonumber\\
&=& e^{-\frac{1}{2}k^2\sigma^2}\cos k \ell
\label{eq:suppress}
\end{eqnarray}

	Therefore, the damping effect of the spatial smearing on the 
oscillating term can be absorbed in the exponential suppression factor of 
Eq.~(\ref{eq:suppress}). For $E=5$ MeV and $\delta m^2$ as high as 
$10^{-9}$ eV$^2$, this suppression factor differs from 1 by only 
$6\times 10^{-5}$. Such a  small  difference is  negligible for our 
purposes. The smearing correction is even smaller than the corrections 
of the order $\varepsilon^2=2.8\times 10^{-4}$, which  can also be 
neglected in the calculation of the near-far asymmetry.




\begin{figure}
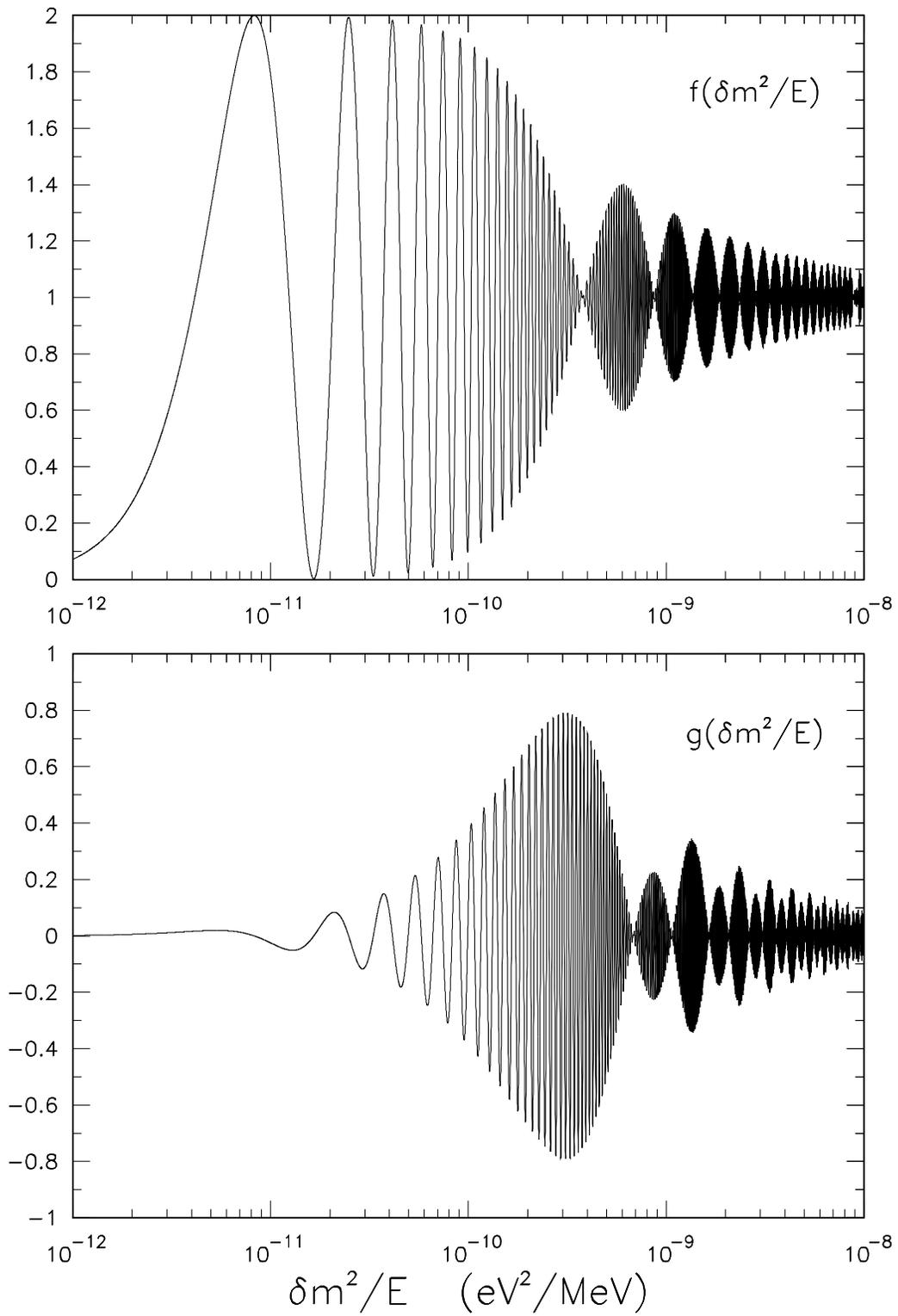

\caption{	The detector-independent functions $f$ and $g$ defined in 
		the text.}
\end{figure}
\begin{figure}
\caption{	SuperKamiokande: Standard electron spectrum $s(T)$ 
		as a function of the measured electron kinetic energy $T$.
		The area under $s(T)$ is normalized to 1. Also shown are the 
		(numbered) energy bins used in the analysis of $A_{NF}$.}
\end{figure}
\begin{figure}
\caption{	SuperKamiokande: Spectra of $\nu_e$ contributing to the 
		entire standard electron spectrum of Fig.~2 or to selected 
		bins. The neutrino spectra are weighted by the interaction 
		cross sections, taking into account the detector threshold 
		and the energy resolution function.}
\end{figure}
\begin{figure}
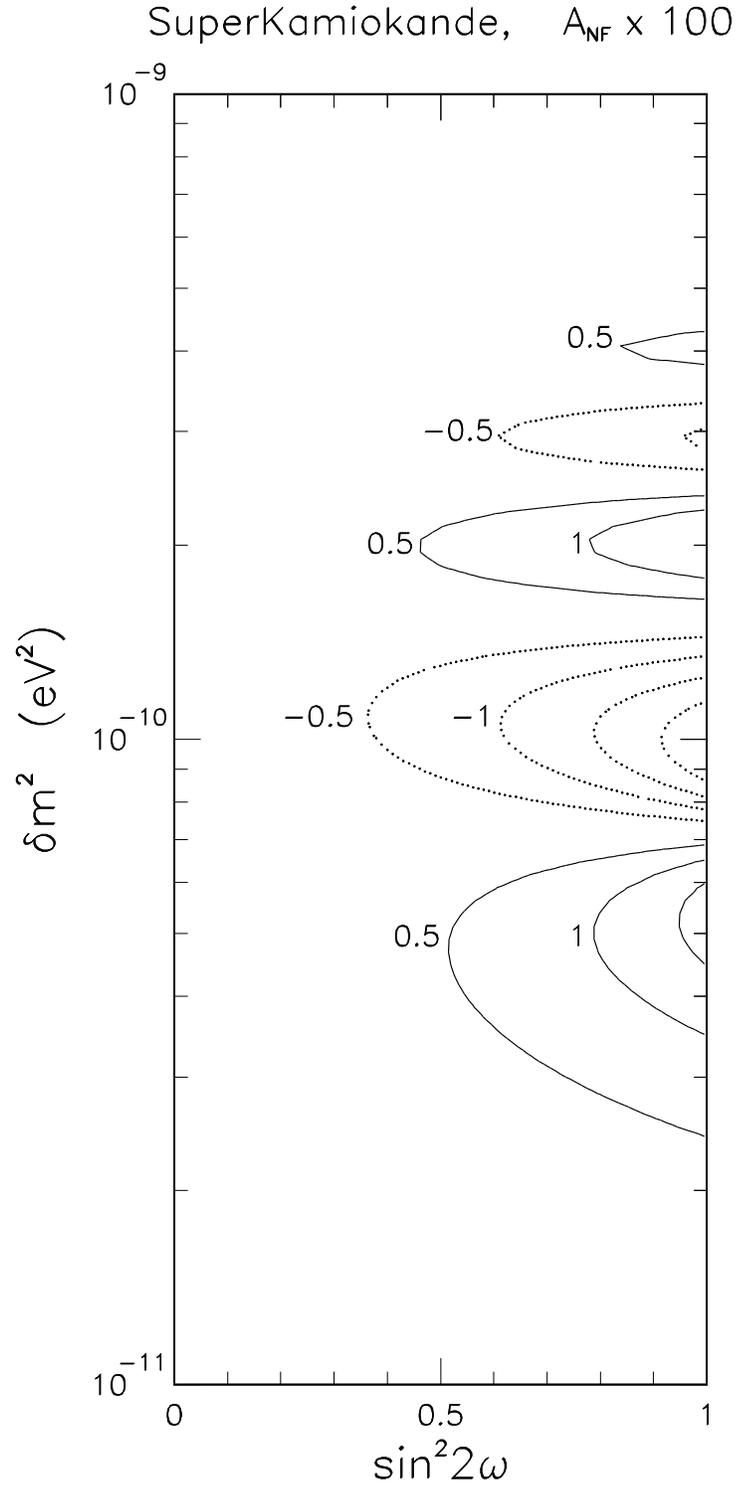

\caption{	SuperKamiokande: Curves of iso-$A_{NF}$ for $2\nu$ 
		oscillations (all bins).}
\end{figure}
\begin{figure}
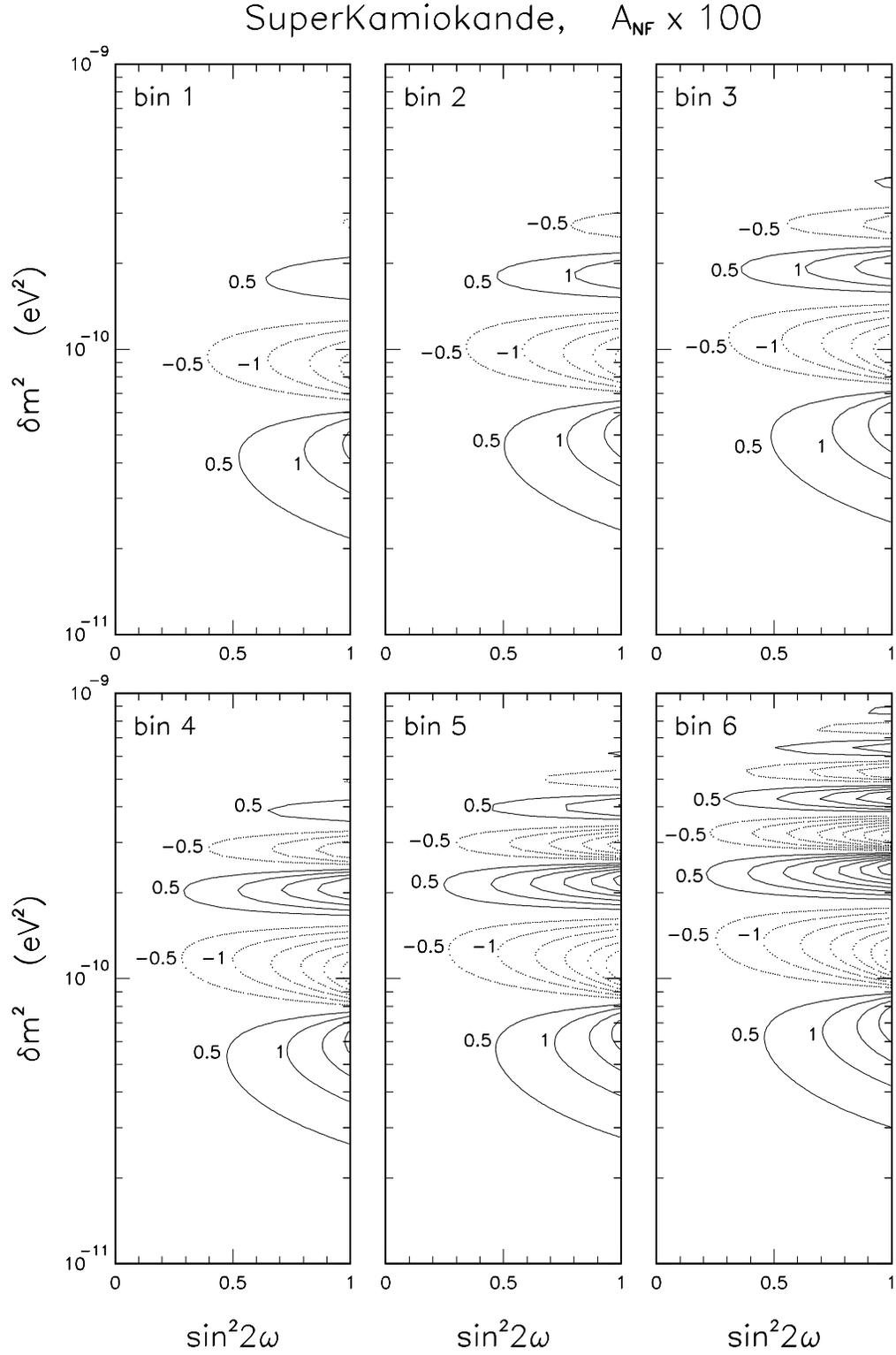

\caption{	SuperKamiokande: Curves of iso-$A_{NF}$  for
		$2\nu$ oscillations (separated bins).}
\end{figure}
\begin{figure}
\caption{	SuperKamiokande: Curves of iso-$A_{NF}$  for
		$3\nu$ oscillations (all bins), for two representative values
		of $\delta m^2$. The triangular representation is discussed 
		in the text.}
\end{figure}
\begin{figure}
\caption{	SNO: Standard electron spectrum $s(T)$  as a function of the
		measured electron kinetic energy $T$.  The area under 
		$s(T)$ is normalized to 1. Also shown are the  (numbered) 
		energy bins used in the analysis of $A_{NF}$.}
\end{figure}
\begin{figure}
\caption{	SNO: Spectra of $\nu_e$ contributing to the entire standard 
		electron spectrum of Fig.~7 or to selected bins. The neutrino
		spectra are weighted by the interaction cross sections, 
		taking into account the detector threshold and the energy 
		resolution function.}
\end{figure}
\begin{figure}
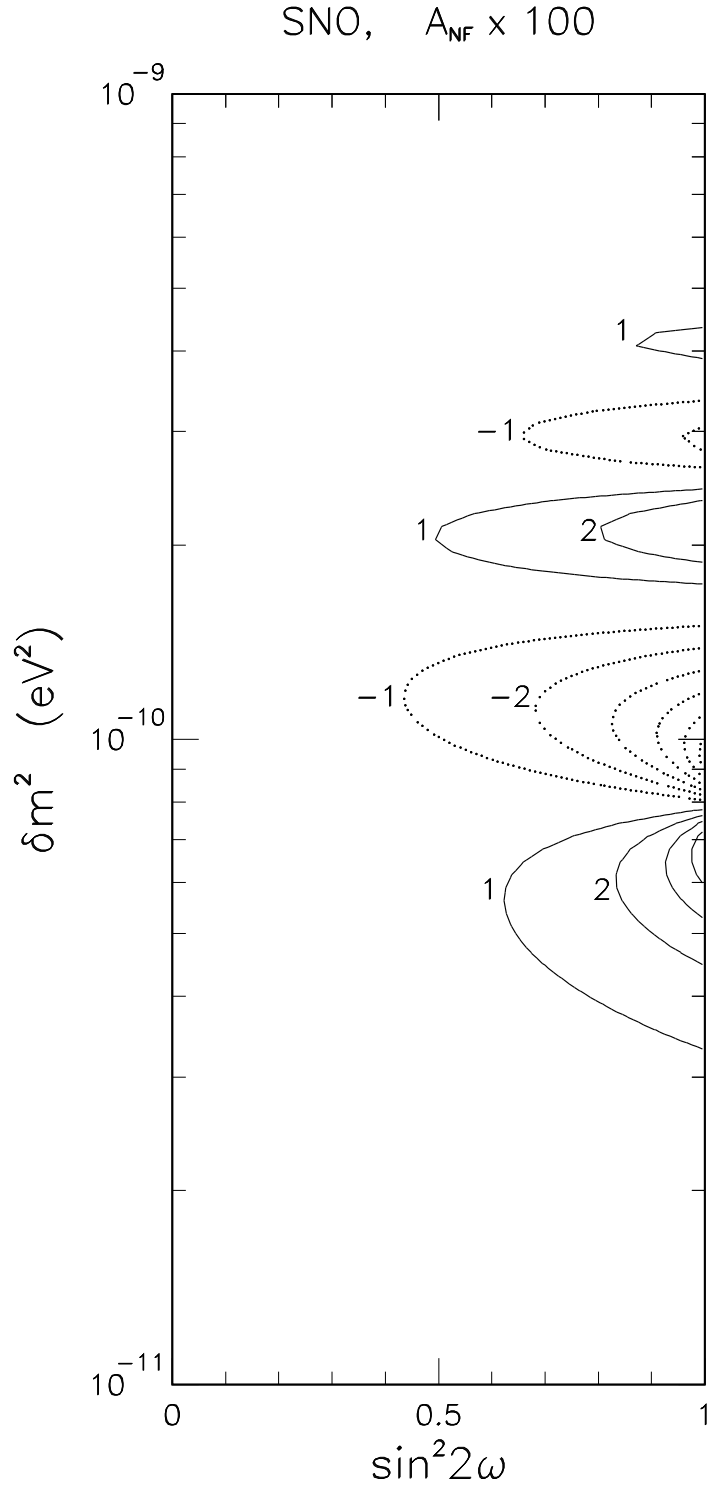

\caption{	SNO: Curves of iso-$A_{NF}$ for  $2\nu$ oscillations 
		(all bins).}
\end{figure}
\begin{figure}
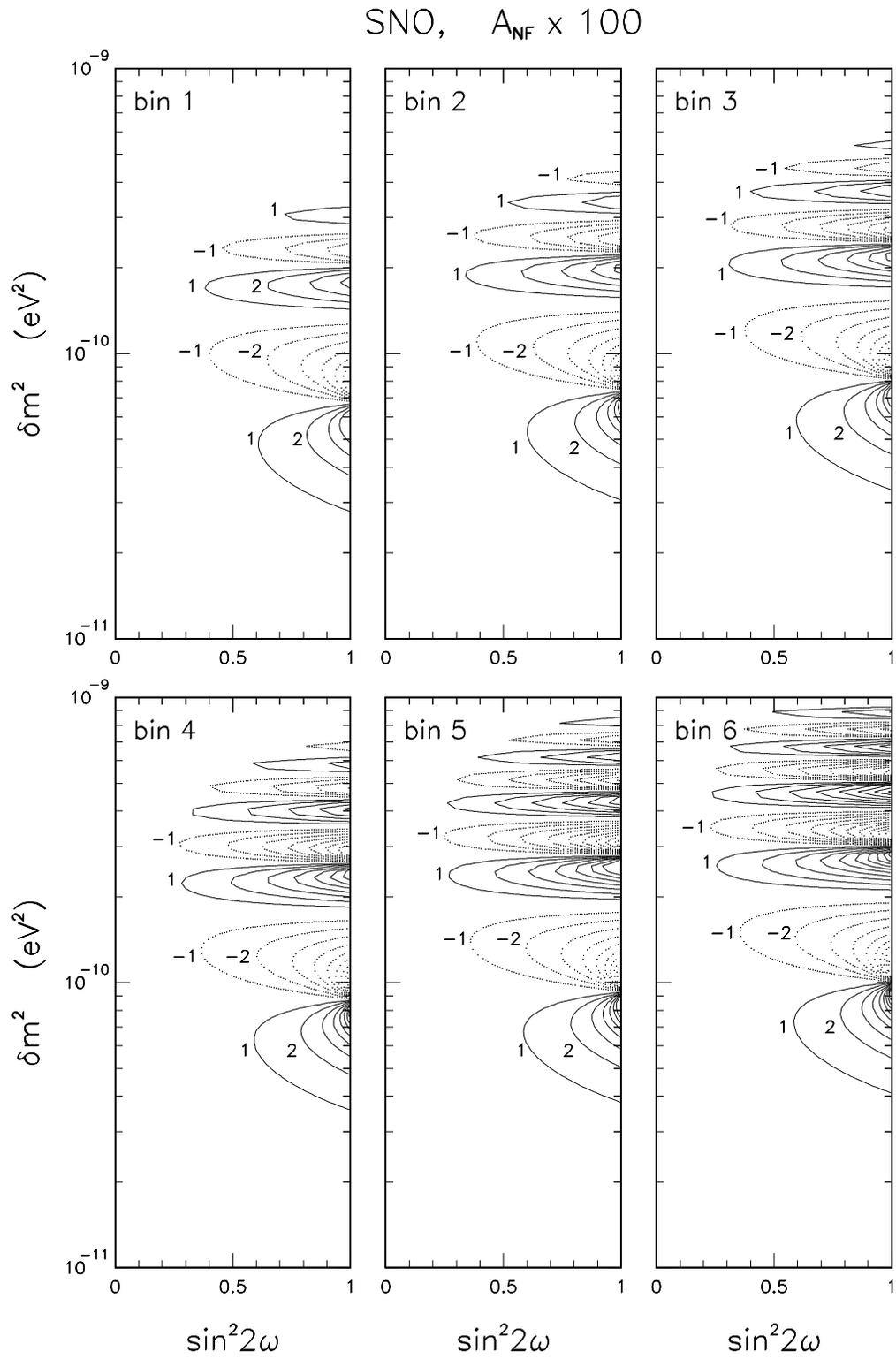

\caption{	SNO: Curves of iso-$A_{NF}$ for  $2\nu$ oscillations 
		(separated bins).}
\end{figure}
\begin{figure}
\caption{	SNO: Curves of iso-$A_{NF}$  for $3\nu$ oscillations 
		(all bins), for two representative values
		of $\delta m^2$. The triangular representation is discussed 
		in the text.}
\end{figure}

\newcommand{\InsertFigure}[2]{\newpage\begin{center}\mbox{%
\epsfig{bbllx=1.4truecm,bblly=1.3truecm,bburx=19.5truecm,bbury=26.5truecm,%
height=22.3truecm,figure=#1}}\end{center}\vspace*{-1.65truecm}%
\parbox[t]{\hsize}{\small\baselineskip=0.5truecm\hskip0.5truecm #2}}
\InsertFigure{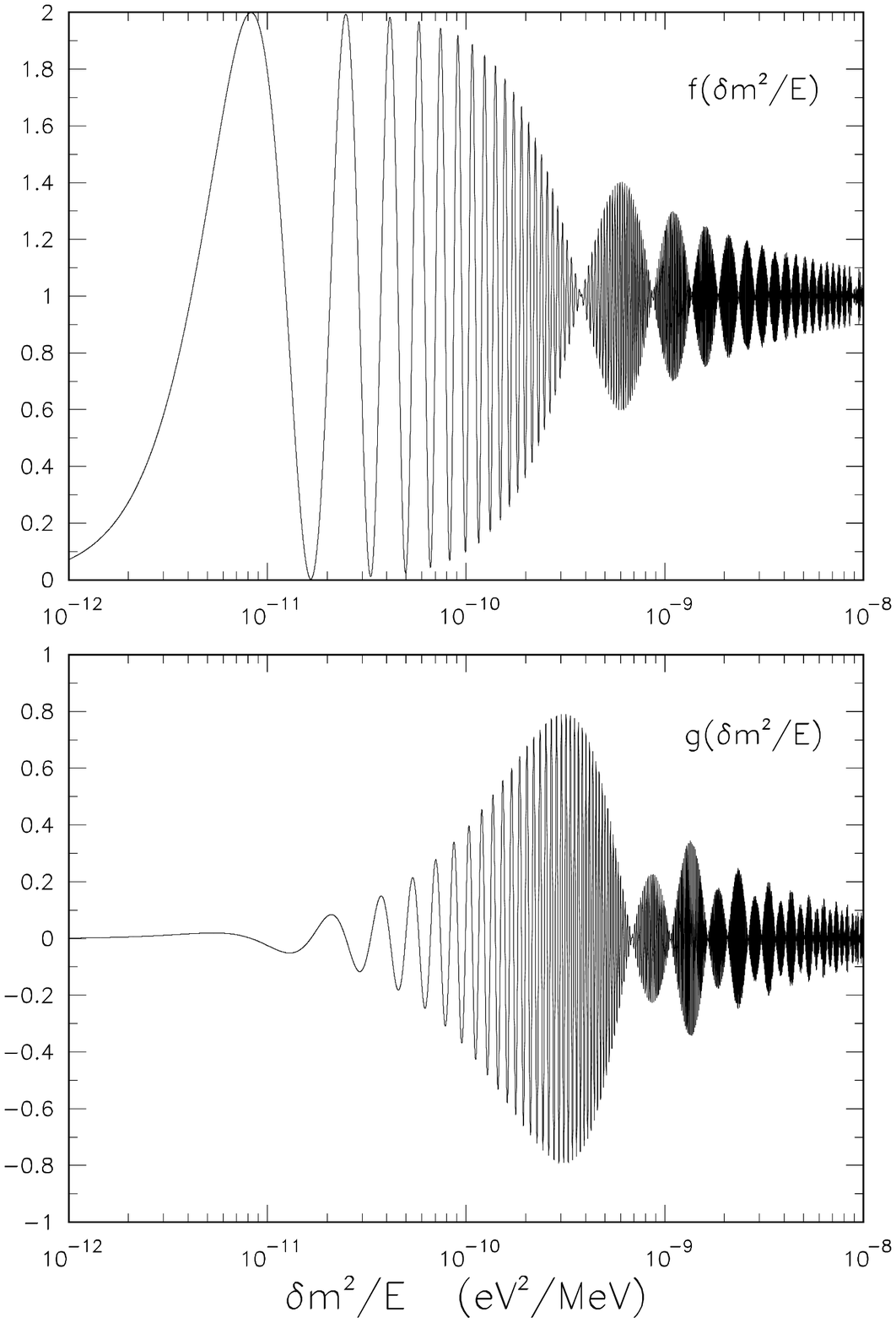}%
{\hfil FIG.~1. 	The detector-independent functions $f$ and $g$ defined in 
		the text.\hfil}
\InsertFigure{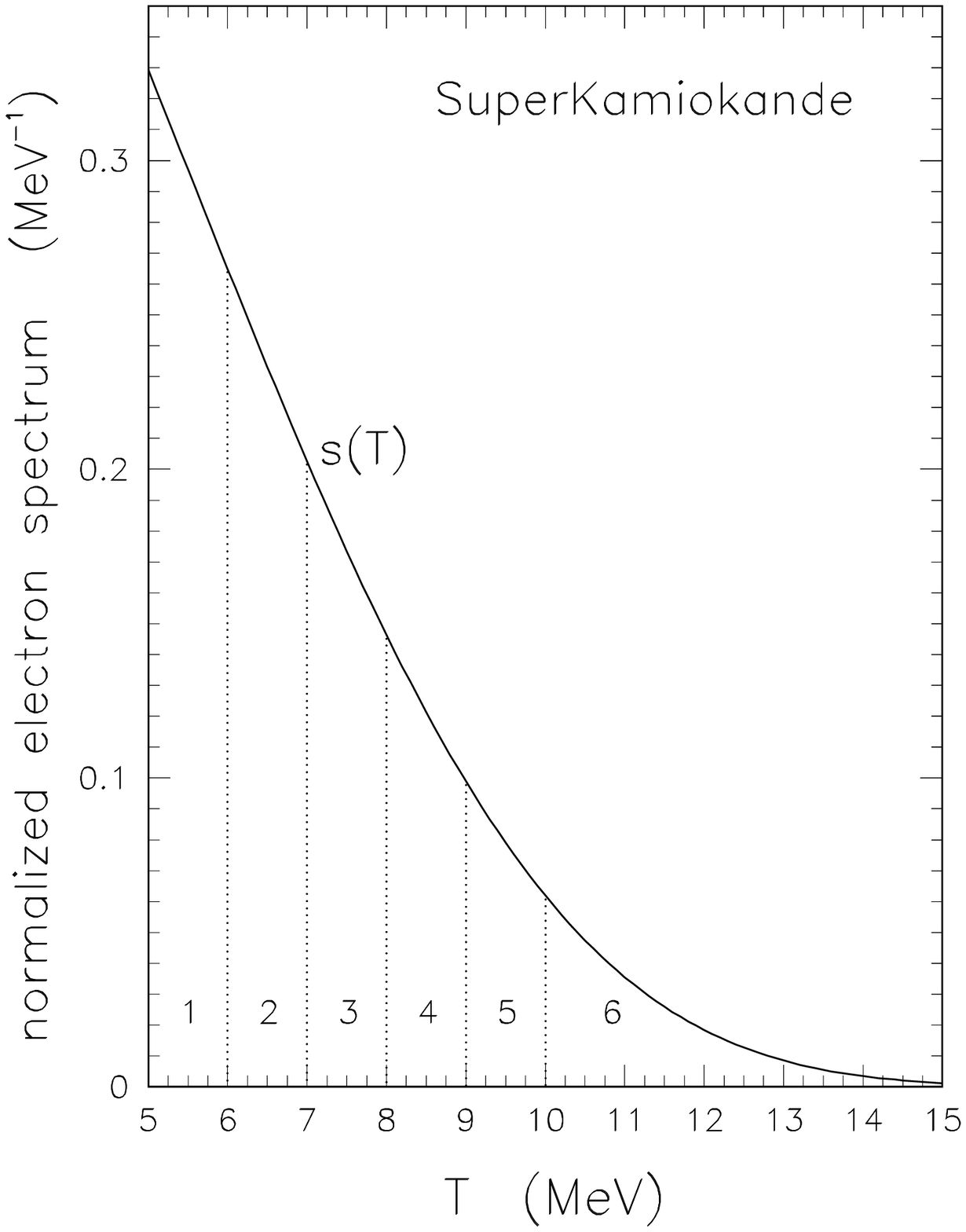}%
{FIG.~2. 	SuperKamiokande:
		Standard electron spectrum $s(T)$ 
		as a function of the
		measured electron kinetic energy $T$.
		The area under $s(T)$ is normalized to 1. Also shown are the 
		(numbered) energy bins
		used in the analysis of $A_{NF}$.}
\InsertFigure{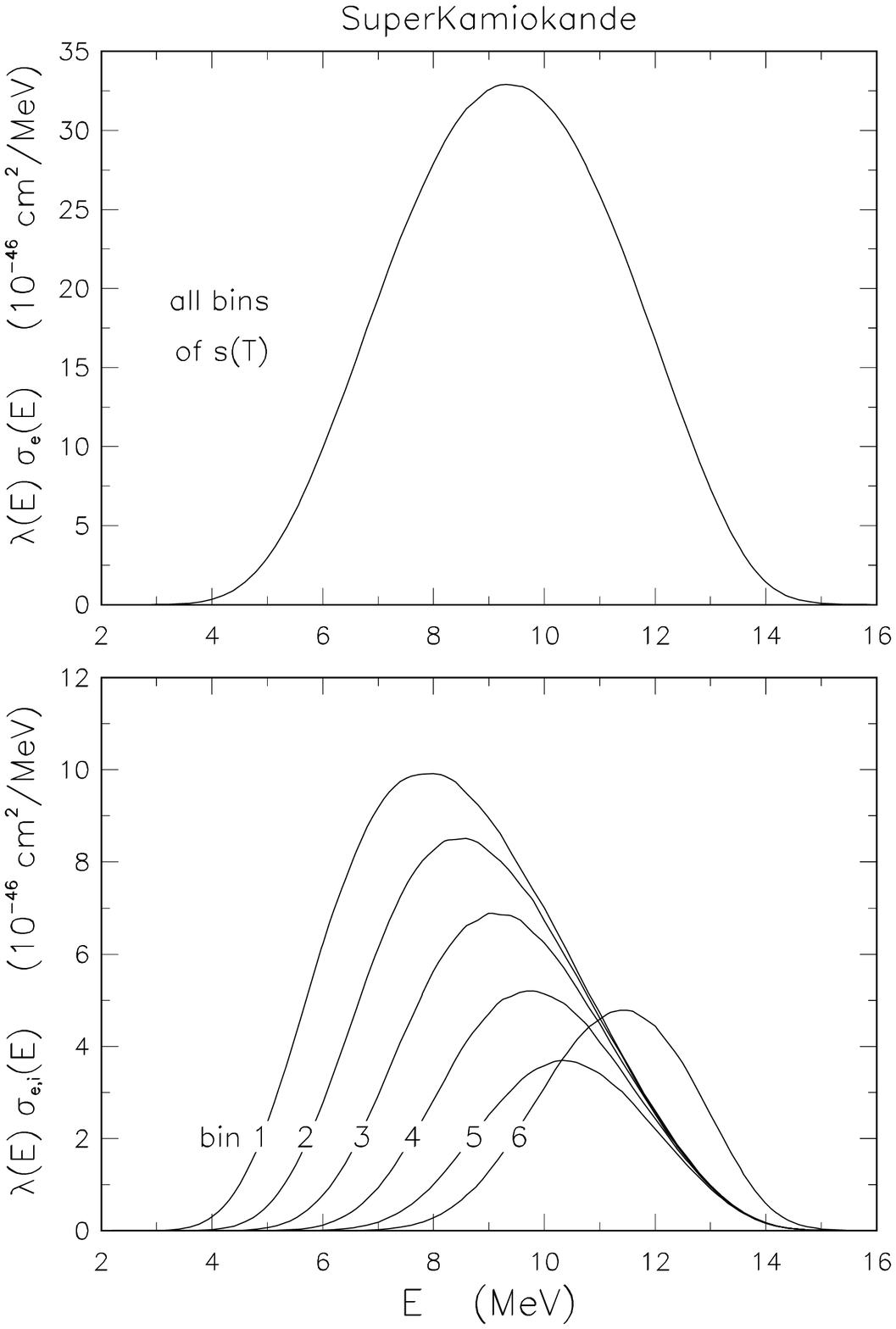}%
{FIG.~3. 	SuperKamiokande:
		Spectra of $\nu_e$ contributing to the entire standard electron
		spectrum of Fig.~2 or to selected bins. The neutrino
		spectra are weighted by the interaction cross sections, 
		taking into account
		the detector threshold and the energy resolution function.}
\InsertFigure{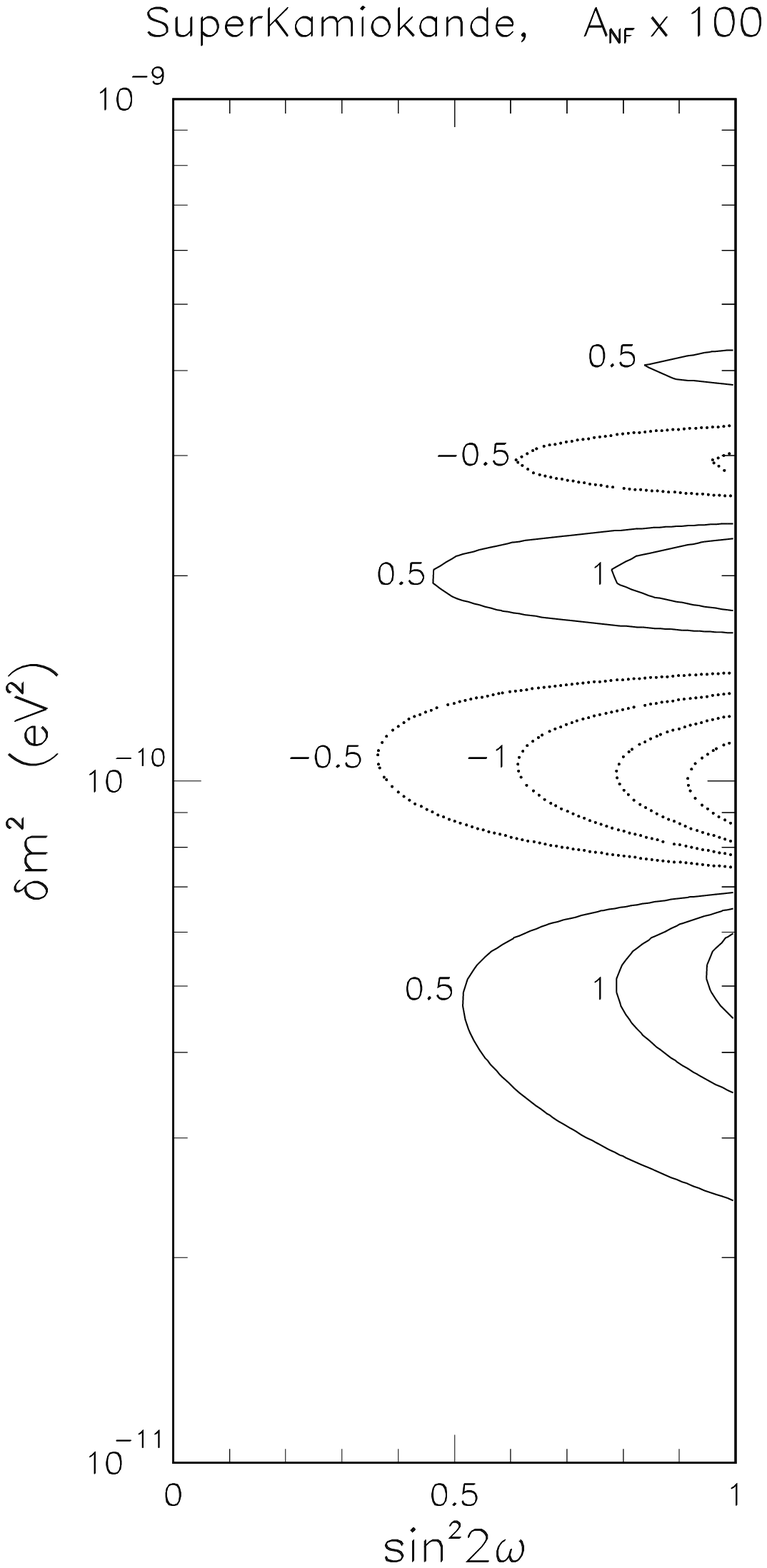}%
{\hfil FIG.~4. 	SuperKamiokande: Curves of iso-$A_{NF}$ for 
		$2\nu$ oscillations (all bins).\hfil}
\InsertFigure{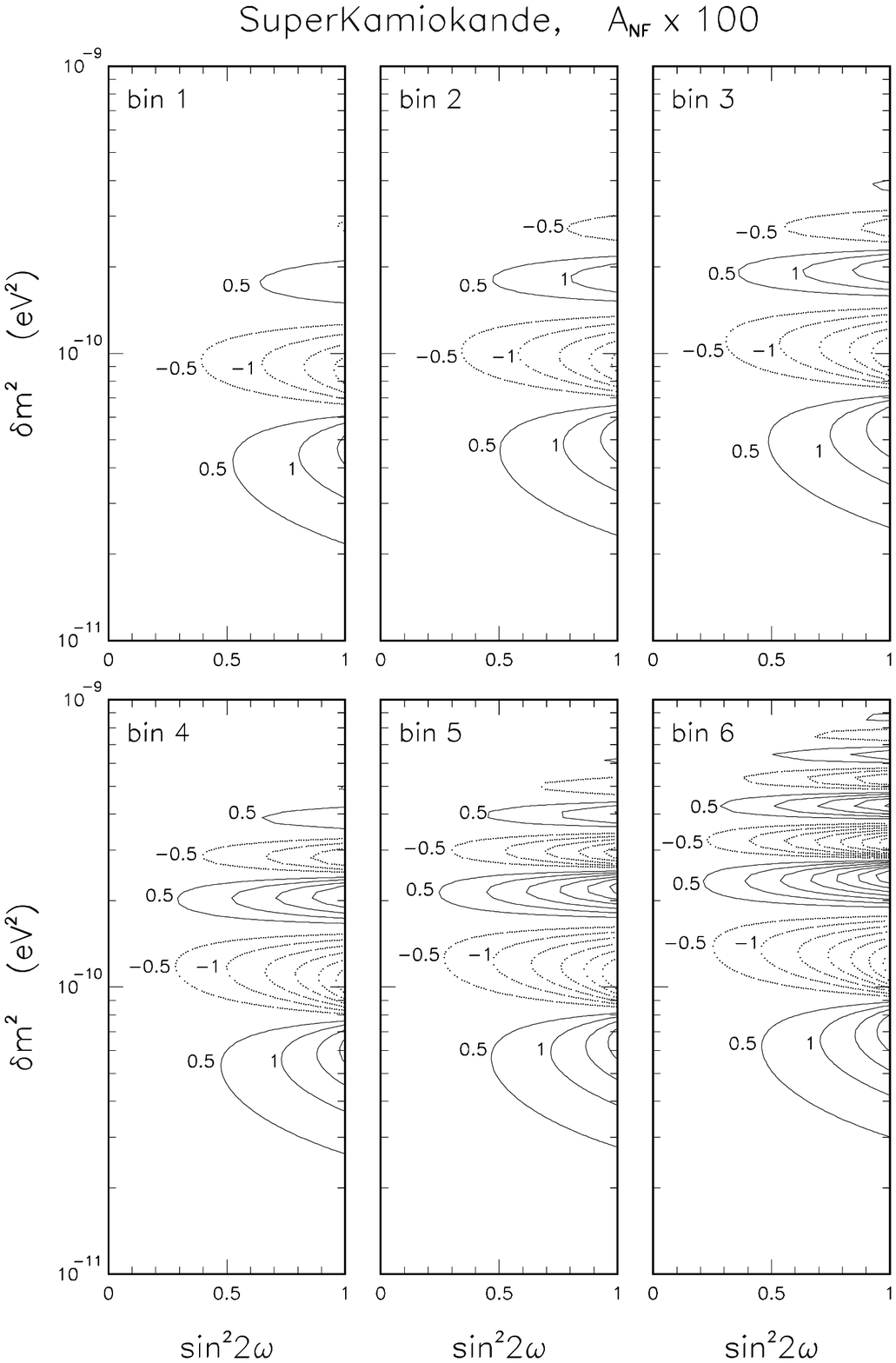}%
{\hfil FIG.~5. 	SuperKamiokande: Curves of iso-$A_{NF}$  for
		$2\nu$ oscillations (separated bins).\hfil}
\InsertFigure{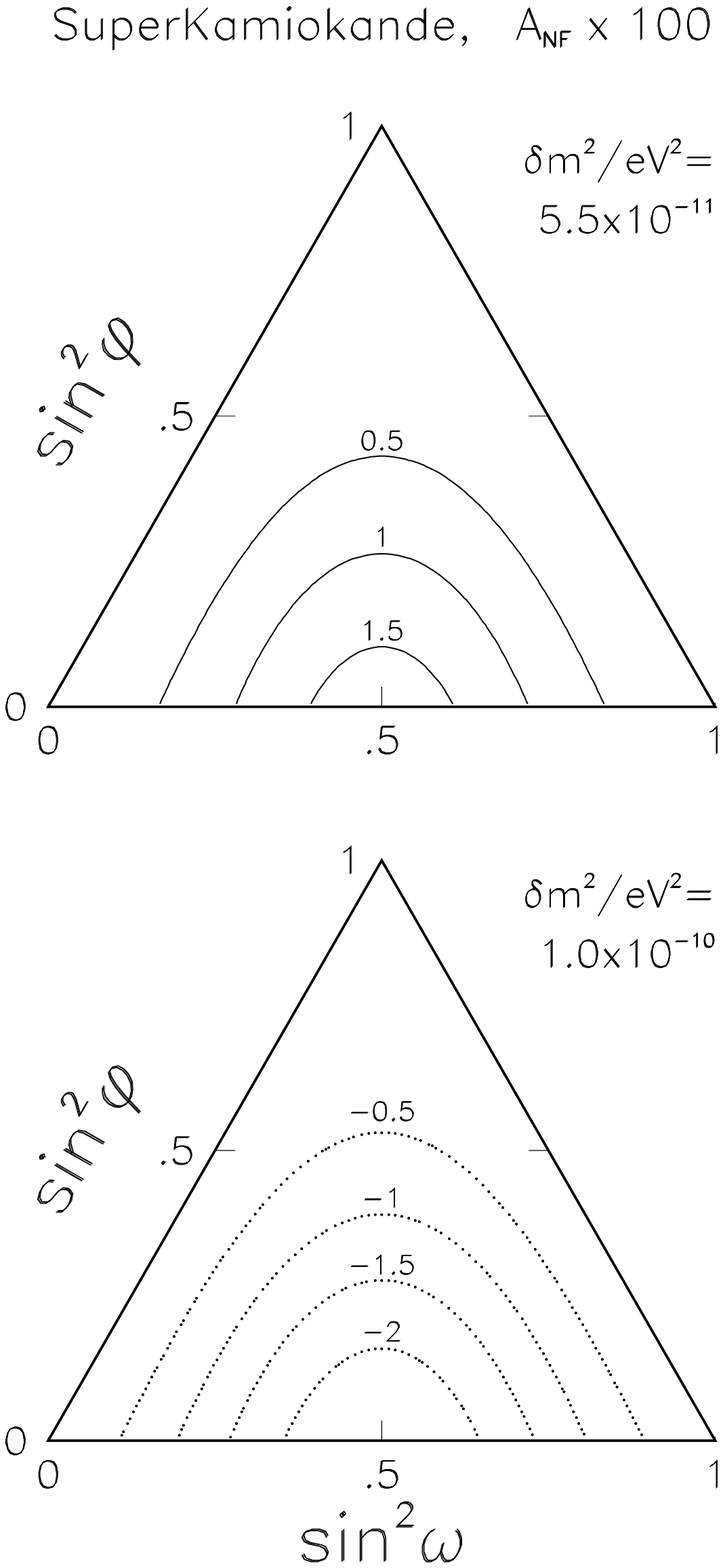}%
{FIG.~6. 	SuperKamiokande: Curves of iso-$A_{NF}$  for
		$3\nu$ oscillations (all bins), for two representative values
		of $\delta m^2$. The triangular representation is 
		discussed in the text.}
\InsertFigure{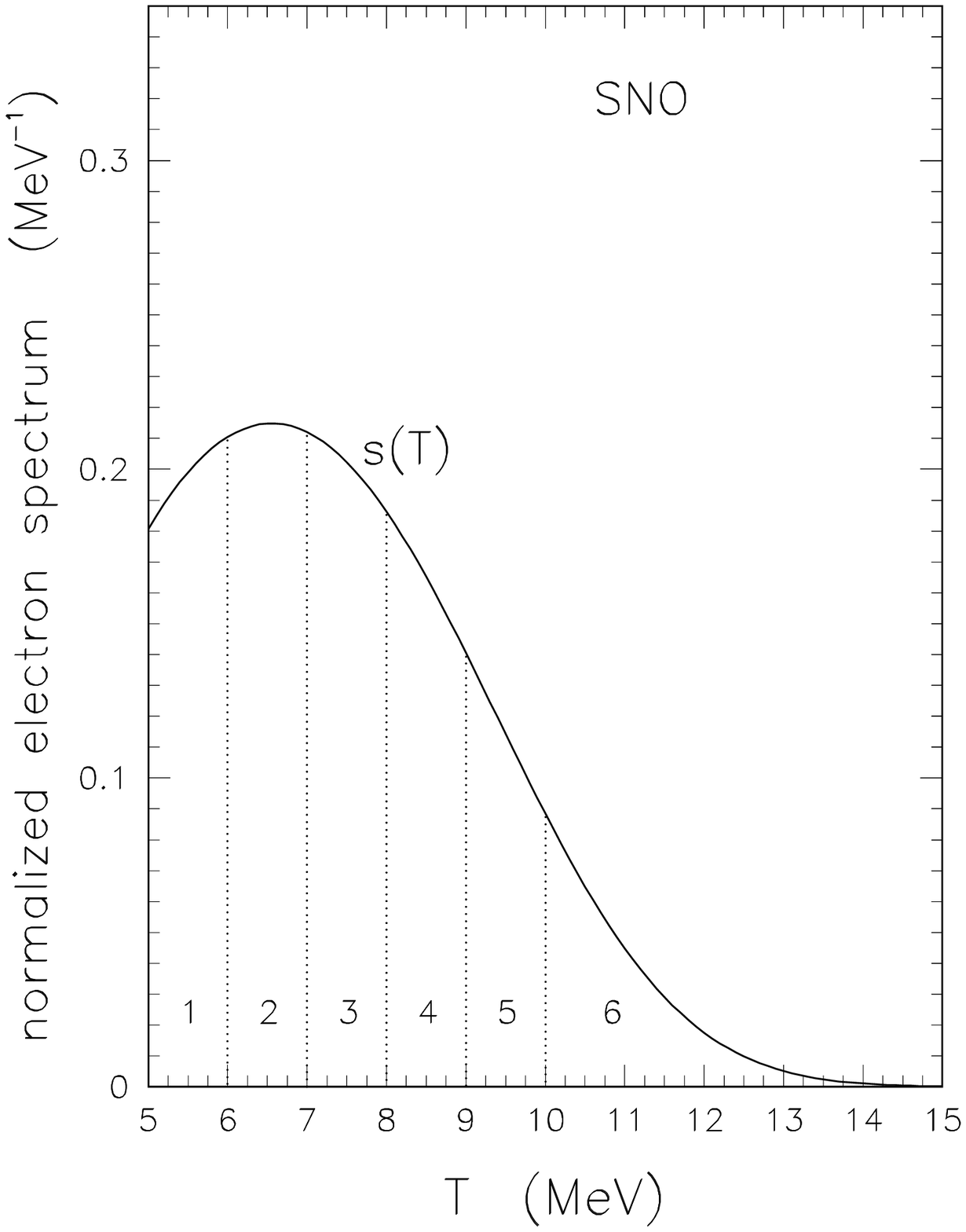}%
{FIG.~7. 	SNO:
		Standard electron spectrum $s(T)$ 
		as a function of the
		measured electron kinetic energy $T$. 
		The area under $s(T)$ is normalized to 1. Also shown are the 
		(numbered) energy bins
		used in the analysis of $A_{NF}$.}
\InsertFigure{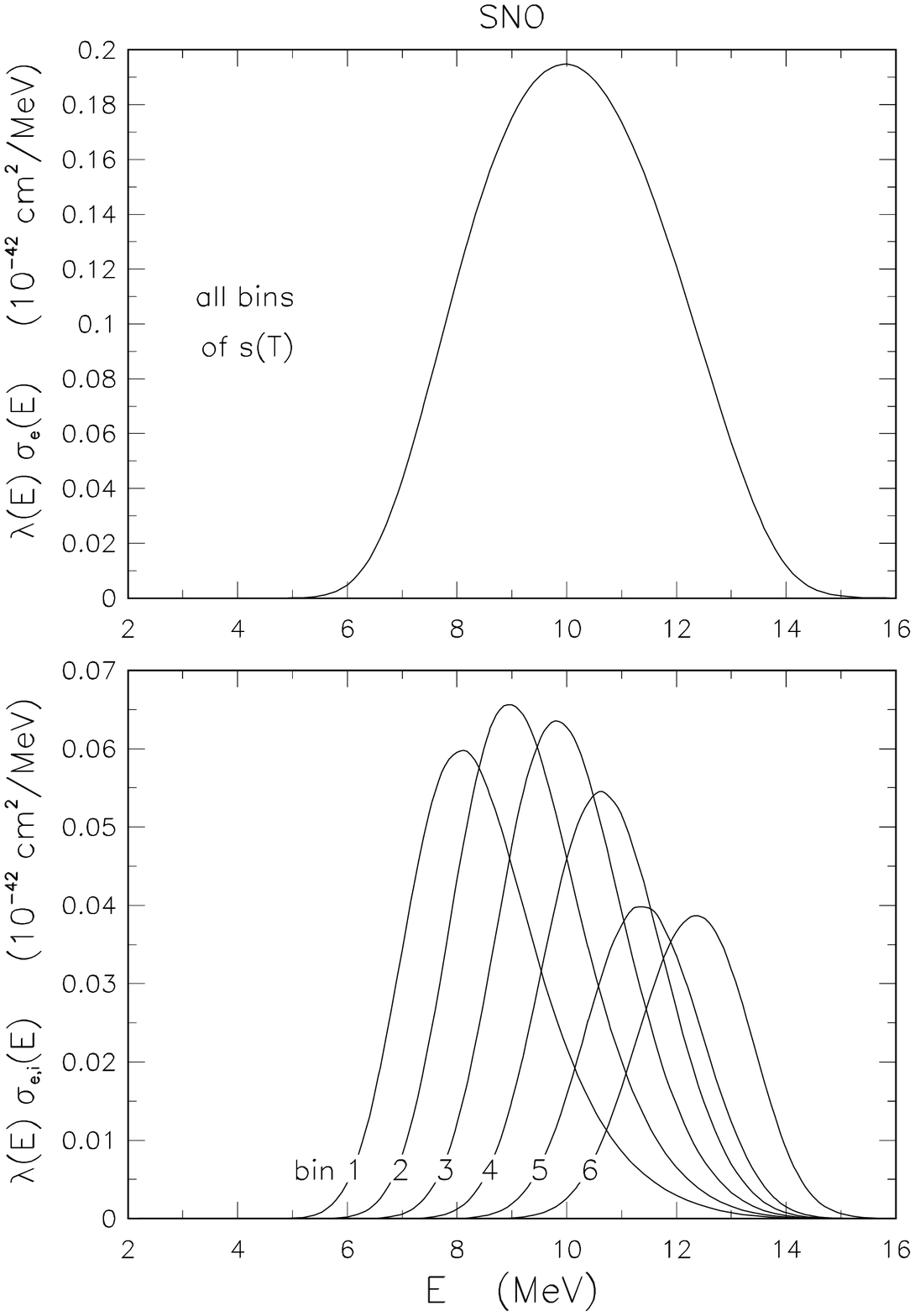}%
{FIG.~8. 	SNO: 
		Spectra of $\nu_e$ contributing to the entire standard electron
		spectrum of Fig.~7 or to selected bins. The neutrino
		spectra are weighted by the interaction cross sections, 
		taking into account
		the detector threshold and the energy resolution function.}
\InsertFigure{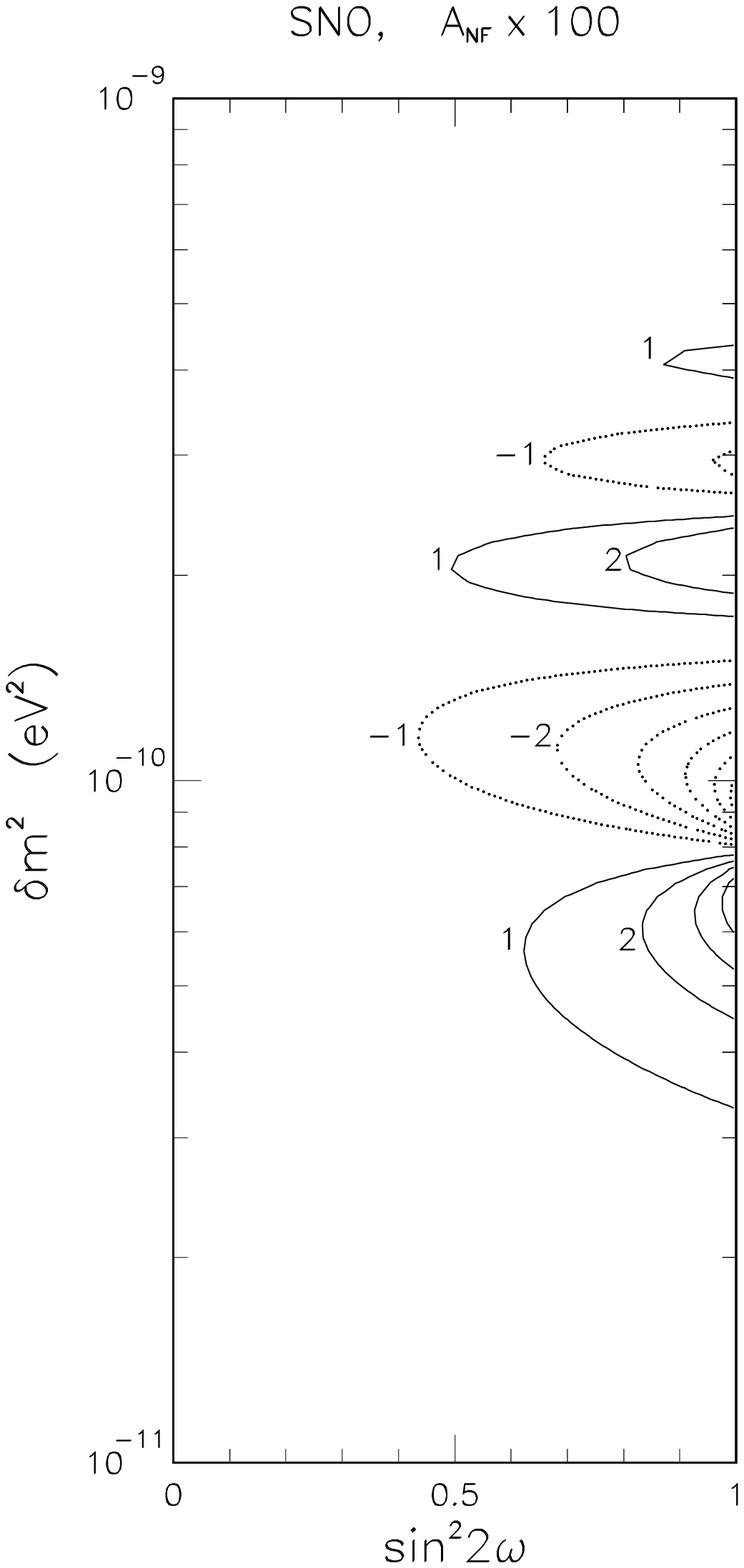}%
{\hfil FIG.~9. 	SNO: Curves of iso-$A_{NF}$ for 
		$2\nu$ oscillations (all bins).\hfil }
\InsertFigure{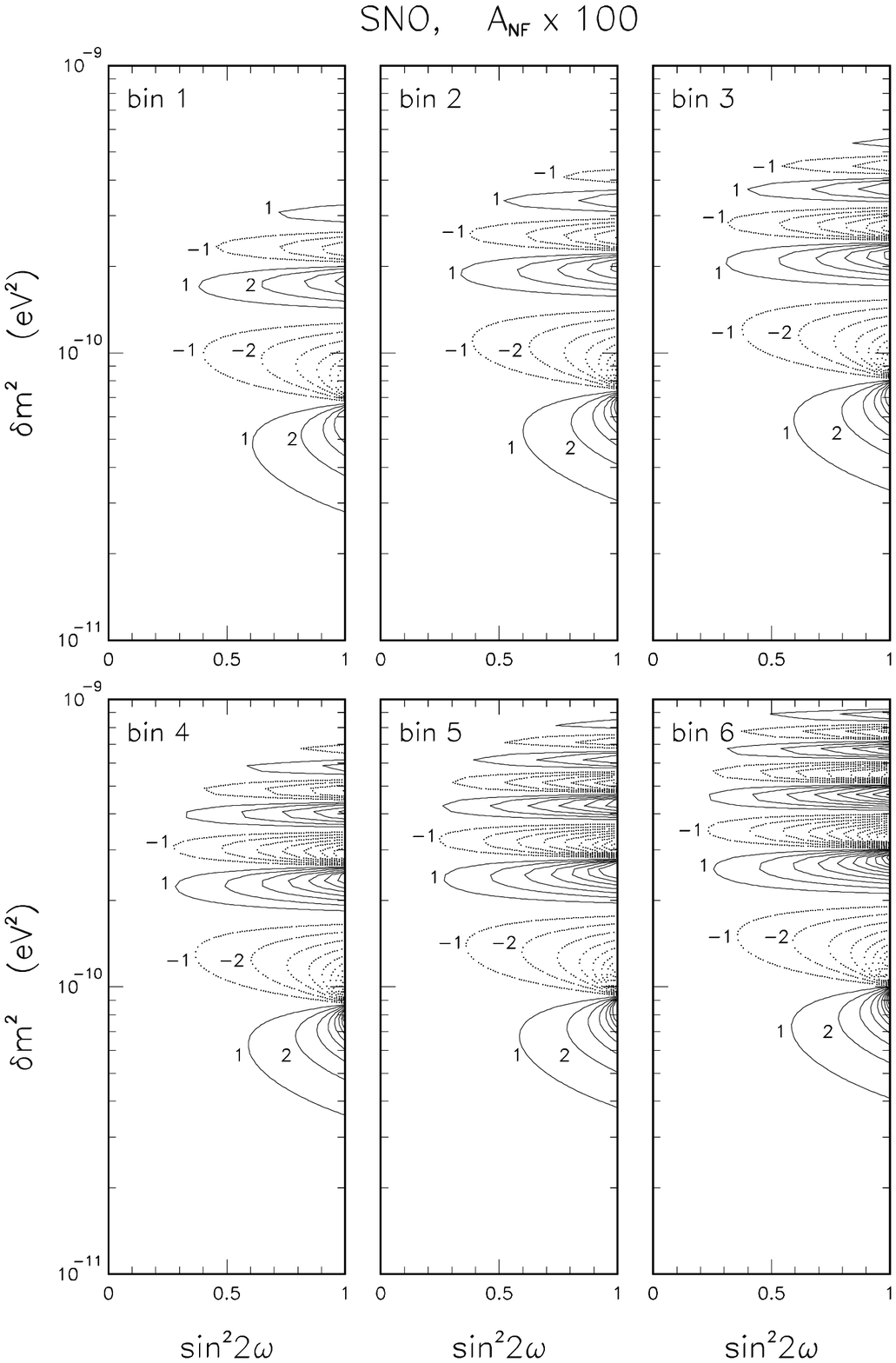}%
{\hfil FIG.~10. SNO: Curves of iso-$A_{NF}$ for  
		$2\nu$ oscillations (separated bins).\hfil}
\InsertFigure{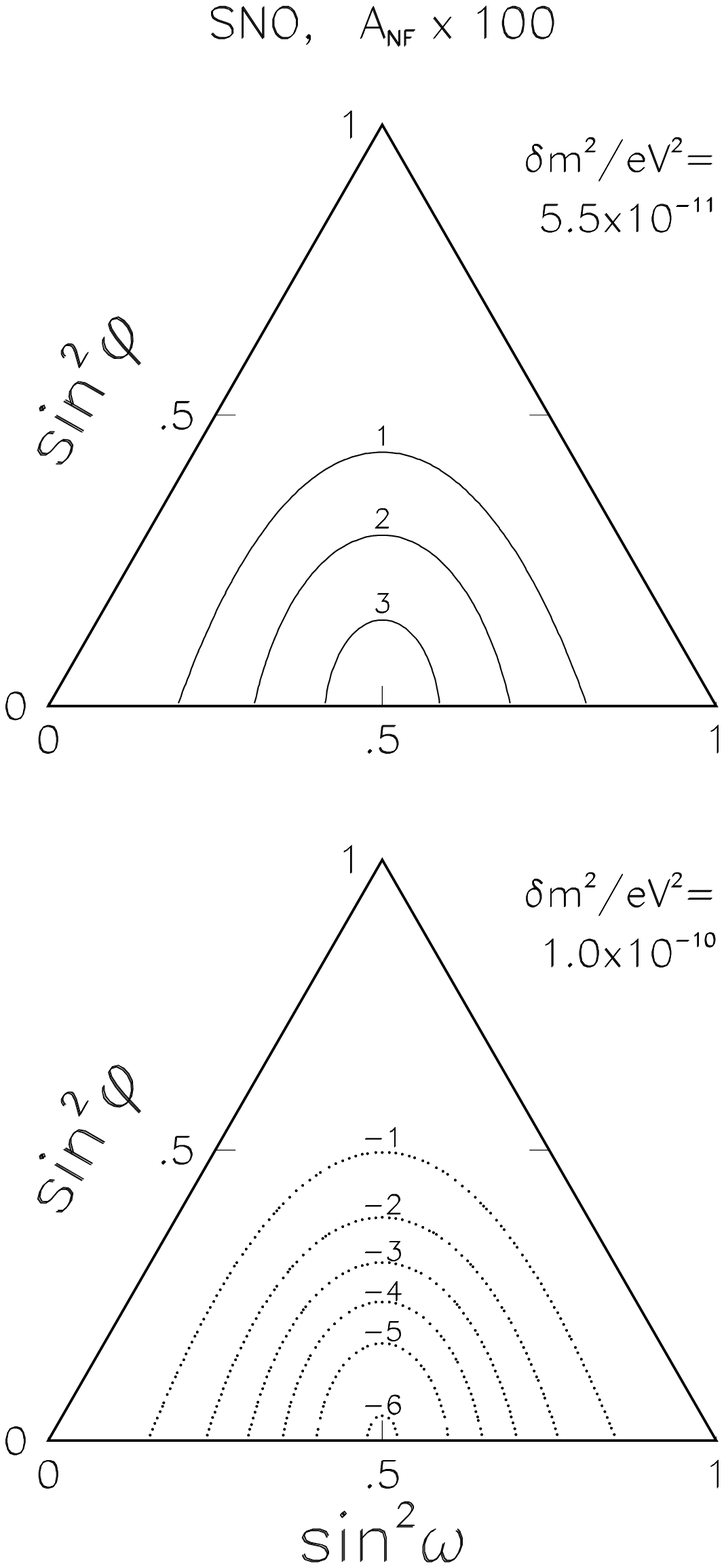}%
{FIG.~11. 	SNO: Curves of iso-$A_{NF}$ for
		$3\nu$ oscillations (all bins), for two representative values
		of $\delta m^2$. The triangular representation 
		is discussed in the text.}

\end{document}